\documentclass[12pt]{article}
\usepackage{amsmath}
\usepackage{setspace}
\usepackage{graphicx}
\usepackage{enumerate}
\usepackage{caption}
\usepackage{natbib,amsthm}
\usepackage{amssymb}
\usepackage{comment}
\newtheorem{theorem}{Theorem}
\newtheorem{lemma}{Lemma}
\newtheorem{corollary}{Corollary}

\newtheorem{remark}{Remark}

\newcommand{\var}{\mathrm{Var}}

\newcommand{\bx}{\boldsymbol{x}}

\newcommand{\bz}{\boldsymbol{z}}
\newcommand{\bs}{\boldsymbol{s}}
\newcommand{\blambda}{\boldsymbol{\lambda}}
\newcommand{\btheta}{\boldsymbol{\theta}}
\newcommand{\beeta}{\boldsymbol{\eta}}
\newcommand{\balpha}{{\mbox{\boldmath$\alpha$}}}
\newcommand{\bbeta}{{\mbox{\boldmath$\beta$}}}
\newcommand{\bzero}{\boldsymbol{0}}
\newcommand{\sigmatt}{{\sigma_r^2}}
\newcommand{\sigmaot}{{\sigma_f^2}}
\newcommand{\bu}{\boldsymbol{u}}
\newcommand{\bc}{\boldsymbol{c}}
\newcommand{\bd}{\boldsymbol{d}}
\newcommand{\bi}{\boldsymbol{i}}
\newcommand{\bI}{\boldsymbol{I}}
\newcommand{\bg}{\boldsymbol{g}}
\newcommand{\bbbdelta}{{\mbox{\boldmath$\delta$}}}

\newcommand{\bmcalI}{\boldsymbol{\mathcal{I}}}
\newcommand{\bxi}{\boldsymbol{x}_i}
\newcommand{\bxoi}{\boldsymbol{z}_{i}}
\newcommand{\ddfrac}[2]{\displaystyle\frac{#1}{#2}}

\newcommand{\removed}[1]{}
\def\T{{ \mathrm{\scriptscriptstyle T} }}

\usepackage{tikz}

\usepackage{A_command}
\usepackage{multirow,xcolor}
\usepackage[figuresright]{rotating}
\title{Survey data integration for regression analysis using model calibration}

\author
{Zhonglei Wang\thanks{Zhonglei Wang is Assistant Professor, Wang Yanan Institute for Studies in Economics (WISE) and School of Economics, Xiamen University, Xiamen, Fujian 361005, PRC (Email: {\tt wangzl@xmu.edu.cn})}
\and
Hang J. Kim\thanks{Hang J. Kim is Associate Professor, Division of Statistics and Data Science, University of Cincinnati, Cincinnati, OH 45221, USA (Email: {\tt kim3h4@ucmail.uc.edu})}
\and
	Jae Kwang Kim\thanks{Jae Kwang Kim is Professor, Department of Statistics, Iowa State University, Ames, IA 50011, USA (Email: {\tt jkim@iastate.edu})}}

\begin{document}
\doublespacing
%  This will produce the submission and review information that appears
%  right after the reference section.  Of course, it will be unknown when
%  you submit your paper, so you can either leave this out or put in 
%  sample dates (these will have no effect on the fate of your paper in the
%  review process!)

%\date{{\it Received XX} XXXX. {\it Revised XX} XXXX.  {\it
%Accepted April} 2007.}
\date{}

\maketitle

%  put the summary for your paper here

\begin{abstract}
We consider regression analysis in the context of data integration. To combine partial information from external sources, we employ the idea of model calibration which introduces a ``working'' reduced model based on the observed covariates. The working reduced model is not necessarily  correctly specified but can be a useful device to incorporate the partial information from the external data. The actual implementation is based on a novel application of the information projection and model calibration weighting. The proposed method is particularly attractive for combining information from several sources with different missing patterns. The proposed method is applied to a real data example combining survey data from Korean National Health and Nutrition Examination Survey and big data from National Health Insurance Sharing Service in Korea. 

\end{abstract}

	\section{Introduction}

Data integration is an emerging research area in survey sampling. By incorporating the partial information from external samples, one can improve the efficiency of the resulting estimator and obtain a more reliable analysis. 
\citet{lohr2017combining},  \citet{yang2020statistical},  and
\citet{rao2020making} provide reviews of statistical
methods of data integration for finite population inference.
Many existing methods \citep[e.g.,][]{hidiroglou2001double,merkouris2010combining,zubizarreta2015stable} are mainly concerned with estimating population means or totals while combining information for analytic inference such as regression analysis is not fully explored in the existing literature. 
% except a few recent researches \citep{sheng2020censored,jiang2021enhanced,sheng2021synthesizing,sheng2021synthesizing2}. 

In this paper, we consider regression analysis in the context of data integration. When we combine data sources to perform a combined regression analysis, we may encounter some problems: covariates may not be fully observed or be subject to measurement errors.  Thus, one may consider the problem as a missing-covariate regression problem. \citet*{robins1994estimation} and \citet*{wang1997} discussed semiparametric estimation in regression analysis with missing covariate data under the missing-at-random covariate assumption. 
%The problem is also related to the meta analysis 
In our setup, the external data source with missing covariate can be a census or big data.

Under this setup, \citet*{chatterjee2016}  developed a data integration method based on the constrained maximum likelihood, which uses a fully parametric model for the likelihood specification and a constraint developed from the reduced model for data integration. The constrained maximum likelihood method is efficient when the model is correctly specified but is not applicable when it is difficult or impossible to specify a correct density function. {\citet*{Kundu2019} generalized the method of \cite{chatterjee2016} to consider multiple regression models based on the theory of generalized
	method of moments \citep[GMM]{hansen1982large}.} Recently, 
\cite{xu2020meta}  develop a data integration method using generalized method of moments technique, but their method implicitly assumes that the reduced model is correctly specified. Under a nested case-control design, \citet*{Shin2020} proposed to use  the fully observed sample in the phase 2 to fit a parametric model, and missing covariates in the phase 1 sample are imputed; also see \citet*{Shin2019}.   \citet*{zhang2021} developed a retrospective empirical likelihood framework to account for sampling bias in case-control studies. \citet*{sheng2021synthesizing2} develop a penalized empirical likelihood approach to incorporate such information in the logistic regression setup.

%We will present the proposed data integration methods first under a simple setting where no selection bias exists in the external data, then under a general setting where the selection bias in the external data is additionally addressed by a probablistic adjustment.

To combine partial information from external sources, we employ the idea of model calibration \citep{wu2001model} which introduces a ``working'' reduced model based on observed covariates. The model parameters in the reduced model are estimated from the  external sources and then combined through a novel application of the empirical likelihood method \citep{owen1991empirical,qin1994}, which can be viewed as information projection \citep{csiszar2004}. 
The working reduced model is not necessarily specified correctly, but a good working model can improve the efficiency of the resulting  analysis. 
The proposed method is particularly attractive for combining information from several data sources with different missing patterns. In this case, we only need to specify different working models for different missing patterns.

Besides, our proposed method is based on the first moment conditions like usual regression analyses, so weak assumptions can broaden the applicability of the proposed method to many practical problems. In particular, the proposed method is directly  applicable to survey sample data which is the main focus of our paper. 
%The computation is simple and straightforward.
% Incorporating extra moment information into estimation is also an important problem in econometrics. 
% {\citet{imbens1994combining} proposed to use an additional moment  incorporating macro averages when analyzing micro data. \citet{hellerstein1999imposing} and \citet{nevo2003using} proposed empirical likelihood methods to adjust the selection probabilities by the available external information; also see \citet{Chaudhuri2008,qin2015using,han2019empirical}.  \citet{Heckman2007} presents a comprehensive summary for data integration in economics.} 
We consider a more general regression setup and 
our proposed empirical likelihood method is different from their empirical likelihood methods and does not require that the working reduced model to be correctly specified. 

We highlight the contribution of our paper as follows. First, we propose a unified framework for  incorporating external data sources in the regression analysis. The proposed method uses weaker assumptions than the parametric model-based method of \cite{chatterjee2016} and thus provides more robust estimation results. Second, the proposed method is widely applicable as it can easily handle multiple external data sources as demonstrated in Section~\ref{sec:MultipleDataIntegration}. 
It can be also applied to the case where the external data source is subject to selection bias.  In the real data application in Section~\ref{sec: application}, we demonstrated that our proposed method can utilize the external big data with unknown selection probabilities by applying propensity score weighting adjustment.  Finally, our proposed method is easy to implement and fully justified theoretically. The computation is simple as it is a direct application of the standard empirical likelihood method and can be easily implemented using the existing software. 
%in Section 4, we have developed a rigorous theory with all the technical details. 

The paper is organized as follows. In Section~\ref{sec:Basic Setup}, a basic setup is introduced, and the existing methods are presented. Section~\ref{sec:Proposed approach} presents the proposed approach, and Section~\ref{sec: theory sec} provides its asymptotic properties. In Section~\ref{sec:MultipleDataIntegration}, an application to multiple data integration is presented. Section~\ref{sec:simulation_study} presents  simulation studies, followed by the application of the proposed method to real data in Section~\ref{sec: application}. Some concluding remarks are made in Section~\ref{sec:conc}. 
	
	\section{Basic Setup} 
\label{sec:Basic Setup}

%\subsection{Basic Setup and Relating Approaches} 

%\textcolor{red}{[Hi Hang, I would like to suggest that we discuss the general method in this section, and put the expectation things as an example or a remark.  ]}

%We first review the semiparametric method of\cite{chatterjee2016}
Consider a finite population $\mathcal{U}=\{1, \ldots, N\}$ of size $N$. Associated with the $i$th unit, let $y_i$ denote the study variable of interest and $\bmx_i=(\bmx_{i1},\bmx_{i2})$ the corresponding auxiliary vector of length $p$. 
% \sout{$\mathcal{U}$ of elements $(\bmx, y)$ which is believed to be an independent and identical realization of random vector $(\bmX, Y)$  with joint density $F(\bmx,y)$ which is completely unspecified}.
% \sout{Without loss of generality, write $\mathcal{U}=\{1, \ldots, N\}$ and  $\bmX=(X_1, X_2)$. Suppose that we} 
We are interested in estimating 
% \sout{parameters} 
a population parameter $\bmbeta_0$, which  solves
% \sout{in a regression model} 
% \begin{equation} 
%  \msout{E( Y \mid x_1, x_2) = m( x_1, x_2 ; \bmbeta)} 
$\bmU_1(\bmbeta) = \sum_{i\in\mathcal{U} }\bmU_1( \bmbeta; \bmx_i, y_i) = \bmzero$ 
%  \label{reg}\notag
%  \end{equation} 
where $\bmU_1( \bmbeta; \bmx, y)$ is a pre-specified estimating function for $\bmbeta$. One example of the estimating function is 
$ \bmU_1( \bmbeta; \bmx_i, y_i) = \{ y_i - m_1(\bmx_i; \bmbeta)\} \bmh_1(\bmx_i; \bmbeta)$, which  is implicitly based on a regression model $E(Y_i \mid \bmx_i) =m_1(\bmx_i; \bmbeta)$ on the super-population level for some $\bmh_1(\bmx_i; \bmbeta)$ satisfying certain identification conditions \citep[e.g.,][]{kim09}. From the finite population a  probability sample $\mathcal{S}_1\subset \mathcal{U}$ is selected,
% \sout{from the finite population}
% \sout{we can construct an   estimating equation for $\bmbeta$  given by} 
and a $Z$-estimator $\hat{\bmbeta}$ can be obtained by solving
\begin{equation} 
	\hat{\bmU}_1 ( \bmbeta) \equiv \sum_{i \in \mathcal{S}_1} d_i \bmU_1( \bmbeta; \bmx_{i}, y_i) = \bmzero,
	\label{eq-1}
\end{equation}
where $d_i$ %$\msout{=\pi_i^{-1}}$ 
is the sampling 
% \sout{design} 
weight for unit $i \in \mathcal{S}_1$.
% \sout{ and 
% $$ U( \bmbeta; \bmx_i, y_i) = \{ y_i - m( \bmx_{i1}, \bmx_{i2}; \bmbeta) \} h(\bmx_{i1}, \bmx_{i2}; \bmbeta)$$
% for some $h (\bmx_{i1}, \bmx_{i2}; \bmbeta)$  such that the solution to \eqref{eq-1} is unique (a.e.)} 
% \sout{Now}

In addition to $\mathcal{S}_1$, suppose that we observe $\bmx_{i1}$ and $y_i$ throughout the finite population and wish to incorporate this extra information to improve the estimation efficiency of $\hat{\bmbeta}$. {Before proposing our method, we introduce two related works, including \citet{chen2000} and \citet{chatterjee2016}.}

%Specifically,  we assume that  an auxiliary vector $\bmalpha^*$ is available, which solves  $\sum_{i\in \mathcal{U}}U_1(\bmalpha;\bmx_{i1},y_i) =\bmzero$ where $\bmx_{i1}$ consists  of $q$ ($<p$) elements of $\bmx_i$.
%In this paper,  we adopt an empirical likelihood framework to incorporate the  auxiliary information. 
\cite{chen2000} first considered this problem in the context of measurement error models. 
To explain their idea in our setup,  we first consider a ``working'' reduced model,  
\begin{equation} 
	E( Y_i \mid \bmx_{i1}) = m_2( \bmx_{i1} ; \bmalpha ) 
	\label{eq-2} 
\end{equation} 
for some $\bmalpha$. 
%Note that since $(\bmx_{i1}, y_i)$ are observed, we can use the population observations to estimate $\hat{\alpha}_N$.  
Under the working model \eqref{eq-2}, we can obtain an estimator $\hat{\bmalpha}$ from the current sample $\mathcal{S}_1$ by solving 
\begin{equation} 
	\hat{\bmU}_2( \bmalpha) \equiv \sum_{i \in \mathcal{S}_1}  d_i \bmU_2( {\bmalpha}; \bmx_{i1}, y_i) = \bmzero,
	\label{eq-2b}
\end{equation} 
where 
$ \bmU_2( {\bmalpha}; \bmx_{i1}, y_i) = \{ y_i - m_2( \bmx_{i1}; \bmalpha) \} \bmh_2(\bmx_{i1}; \bmalpha)$ for some  $\bmh_2(\bmx_{i1}; \bmalpha)$ satisfying conditions similar to ones imposed to $\bmh_1(\bmx_i;\bmbeta)$. 
%$\bmalpha = \bmalpha^*$. 
{In addition, one can get ${\bmalpha}^*$ that solves 
$\sum_{i =1}^N \bmU_2( {\bmalpha}; \bmx_{i1}, y_i) = \bmzero$.}  \cite{chen2000}  proposed using 
% \begin{equation} 
$$\hat{\bmbeta}^* = \hat{\bmbeta} + \widehat{Cov} ( \hat{\bmbeta}, \hat{\bmalpha} ) \{ \hat{V} ( \hat{\bmalpha} ) \}^{-1}   \left({\bmalpha}^* - \hat{\bmalpha} \right) $$
%  \label{eq-3}
%  \end{equation}
as an efficient estimator of $\bmbeta$ 
where $\hat{V} (\cdot)$ and $\widehat{Cov} (\cdot)$ denote the design-based variance and covariance estimators, respectively.   The working model in \eqref{eq-2} is  not necessarily correctly specified, but a good working model can improve the efficiency of the final estimator. While the estimator of   \citet{chen2000} is theoretically justified, it can be numerically unstable as the estimation errors of the variance and covariance matrix can be large. 

%This is also  the main idea of model calibration in \cite{wu2001model}. 
%\textcolor{red}{(Note that we use $U_1$ to denote the estimating function for $\alpha$.)} \zl{(Yes, we have used a subscript ``1'' to denote the functions for $\bmbeta$ and ``2'' for ``$\bmalpha$''.)}

\cite{chatterjee2016} considered a likelihood-based approach using a conditional distribution of $Y_i$ given $\bmX_i$ with density $f( y_i \mid \bmx_i; \bmbeta)$ and imposed a constraint {based on external information. Specifically, they proposed to maximize }
{
\begin{equation}
    \prod_{i\in\mathcal{S}_1}f( y_i \mid \bmx_i; \bmbeta)d F(\bmx_i)
\end{equation}}
subject to
\begin{equation} 
	\int\int \bmU_2( \bmalpha^*; \bmx_{1}, y) f (y\mid \bmx ; \bmbeta) d y {d F(\bmx)}= \bmzero,
	\label{Chatterjee_const} 
\end{equation} 
{where $F(\bmx)$ is an unspecified distribution function for $\bmx$, $dF(\bmx)$ is the Radon-Nikodym derivative of the distribution function $F(\bmx)$ with respect to a certain dominating measure}, and $\bmalpha^*$ is the model parameter available from an external source. 
Following the likelihood based approach of  \citet{chatterjee2016}, $\bmU_2( \bmalpha; \bmx_{i1}, y)$ corresponds to the estimating function involving a ``reduced'' distribution function $g(y_i\mid \bmx_{i1};\bmalpha_0)$ with model parameter $\bmalpha_0$, where $g(y_i\mid \bmx_{i1};\alpha_0)$ can be incorrectly specified. That is, $\bmalpha^*$ is the external information for $\bmalpha_0$. \citet{chatterjee2016} estimated $F(\bmx)$ nonparametrically by empirical likelihood. 
%Constraint \eqref{Chatterjee_const} can be understood as a constraint for the parameter $\bmbeta$ with constraint  $ E\{ \bmU_2 ( \bmalpha^* ; \bmx_{i1}, Y_i ) \mid \bmx_i; \bmbeta \} = \bmzero. $ 
By imposing this constraint into the maximum likelihood estimation, the external information $\bmalpha^*$ can be naturally incorporated.

The constrained maximum likelihood (CML) method is not directly applicable to our conditional mean model in \eqref{eq-1} as the likelihood function for $\bmbeta$ is not defined in our setup. {Besides, the design feature for the probability sample $\mathcal{S}_1$ is not directly applicable in their method.} Nonetheless, one can use an objective function such as that in generalized method of moments to apply the constrained optimization problem, which is asymptotically equivalent to the empirical likelihood method \citep{imbens2002}.  
%\cite{chatterjee2016} also noted that the CML approach could be formulated using the empirical likelihood method of \cite{qin1994} and \cite{qin2000}. 
The empirical likelihood implementation of CML approach is discussed by \cite{han2019empirical}.

	\section{Proposed Approach} % {Proposed method} 

\label{sec:Proposed approach}

We now consider an alternative approach for combining information from several sources. To combine information from several sources, we use the KL divergence measure to apply the information projection  \citep{csiszar2004} on the  model space with constraints. Let  ${\hat{P}}$ be the empirical distribution of the sample {with }
{
	\begin{equation} 
		{\hat{P}} (x, y) = \frac{1}{ \sum_{i \in \mathcal{S}_1} d_i }  \sum_{i \in \mathcal{S}_1} d_i   \mathbb{I} \{ (x,y) = {(x_i, y_i)} \}.
		\label{emp}
	\end{equation} 
}Given the empirical distribution $\hat{P}$, we wish to find the minimizer of 
\begin{equation} 
	D({\hat{P}} \parallel {P})
	= \int \log \{ d  {\hat{P}} (\bmx,y) \}  d   {\hat{P}} (\bmx,y)- \int \log\{ d {P}  (\bmx,y) \} d {\hat{P}} (\bmx,y) 
	\label{divergence}
\end{equation} 
with respect to $P$ in the model space. Notice that the first term is a constant and 
the minimizer of (\ref{divergence}) is the pseudo maximum likelihood estimator of $\hat{P}$.  

{We consider the following constraints in our model at the finite-population level:}  
\begin{equation} 
	\sum_{i=1}^N \bmU_1( \bmbeta; \bmx_i, y_i)  p(\bmx_i, y_i) = 0 \  \mbox{ and } \  \sum_{i=1}^N  \bmU_2( \bmalpha^*; \bmx_{i1}, y_i) p(\bmx_{i}, y_i) = 0 , 
	\label{constraint2}
\end{equation} 
where  $p(\bmx_i, y_i)$ is the point mass assigned to point $(\bmx_i, y_i)$ in the finite population  {satisfying}  $\sum_{i=1}^N p(\bmx_i, y_i)=1$.  See Figure 1 for a graphical illustration of the information projection. 

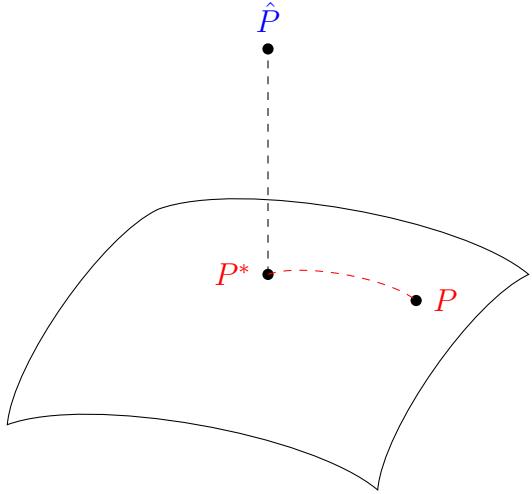
\begin{figure}
	\begin{center}
		
		\begin{tikzpicture}[x={(170:1cm)},y={(55:.7cm)},z={(90:1cm)}]
			\draw[solid,looseness=.6] (2.5,-2.5,-1) node[above right] {}
			to[bend left] (2.5,2.5,-1)
			to[bend left] coordinate (mp) (-2.5,2.5,-1)
			to[bend right] (-2.5,-2.5,-1)
			to[bend right] coordinate (mm) (2.5,-2.5,-1)
			-- cycle;
			\node [fill=black, shape=circle, inner sep=1.5pt, label={[blue]above:$\hat P$}] (O) at (0,0,3) {};
			\node [fill=black, shape=circle, inner sep=1.5pt, label={[red]left:$P^*$}] (O) at (0,0,0) {};
			\node [fill=black, shape=circle, inner sep=1.5pt, label={[red]right:$P$}] (O) at (-2,0,0) {};
			\draw[dashed] (0,0,0) -- (0,0,3);
			\draw[red, dashed, looseness=.6] (0,0,0) to[bend left] (-2,0,0);
			
		\end{tikzpicture}
		\caption{Information projection for the empirical distribution $\hat{P}$. Note that $P^*$ minimizes $D( \hat{P} \parallel P)$ among $P$ satisfying the constraints in (\ref{constraint2}).} 
	\end{center} 
\end{figure}

Using the weighted empirical distribution in  (\ref{emp}),  the KL divergence measure in (\ref{divergence}) reduces to 
$
	D({\hat{P}} \parallel {P})
	=  \mbox{constant} - \hat{N}^{-1} \sum_{i \in \mathcal{S}_1} d_i \log\{p (\bmx_i,y_i) \} 
$ 
where $\hat{N}= \sum_{i \in S_1} d_i$.  
Thus, we only have  to maximize $l ( \mathbf{p}) = \sum_{i \in \mathcal{S}_1} d_i \log (p_i) $ 
subject to $\sum_{i=1}^N {p_i} =1$ and the constraints in (\ref{constraint2}), {where $p_i$ abbreviates $p(\bmx_i,y_i)$}. Note that having  $p_i>0$ for $i \notin \mathcal{S}_1$ will decrease the value of  $l ( \mathbf{p}) = \sum_{i \in \mathcal{S}_1} d_i \log (p_i) $,   the solution ${\hat{p}_i}$ to this  optimization problem 
should give $\hat{p}_i=0$ for $i \notin S_1$.  
Therefore, we can safely set $p_i=0$ for $i \notin S_1$ and 
express the problem as finding the maximizer of 
\begin{equation} 
	Q(\bmd, \bmw) = \sum_{i \in \mathcal{S}_1} d_i \log (w_i) \label{eq: Q(d,w)}
\end{equation}
subject to 
\begin{eqnarray} 
&\displaystyle\sum_{i \in \mathcal{S}_1}  w_i =1,\label{eq: sum to 1} \\ 
&\displaystyle	\sum_{i \in \mathcal{S}_1}  w_i \bmU_2( {\bmalpha}^*; \bmx_{i1}, y_i) = \bmzero,
	\label{eq-4}\\
&\displaystyle	\sum_{i \in \mathcal{S}_1} {w}_i \bmU_1( \bmbeta; \bmx_{i}, y_i)  = \bmzero. \notag
\end{eqnarray}
We use $w_i$ instead of $p_i$ to represent the final weights assigned to the  sample elements.

\begin{remark}
Maximizing the objective function in (\ref{eq: Q(d,w)})  is equivalent to minimizing the following cross entropy:
\begin{equation}
-\sum_{i\in \mathcal{S}_1}\tilde{d}_i\log(w_i),\label{eq: rem cross entropy}
\end{equation}
where $\tilde{d}_i=d_i/(\sum_{i\in \mathcal{S}_1}d_i)$. The objective function (\ref{eq: rem cross entropy}) is also 
the pseudo empirical log-likelihood function 
considered by \citet{chensitter1999} and \citet{wu2006}. Instead of (\ref{eq: Q(d,w)}),  we may consider other objective functions, including the  population empirical likelihood proposed by  \citet{ChenKim2014} for example.
\end{remark}

Our proposed method is different from \cite{chatterjee2016} in that we use a more general integral constraint
\eqref{Chatterjee_const} which does not involve the conditional density function $f( y \mid \bmx; \bmbeta)$. 
Constraint (\ref{eq-4}) still  incorporates the extra information in $\bmalpha^*$.  
The above optimization can be solved by applying the standard profile empirical likelihood method or using the following two-step estimation method. 
\begin{enumerate} 
	
	\item Find the calibration weights $\hat{\bmw} = \{\hat{w}_i : i \in \mathcal{S}_1\}$ maximizing $Q(\bmd, \bmw)$ subject to \eqref{eq: sum to 1}--\eqref{eq-4}. 
	\item Once the solution $\hat{\bmw}$ is obtained from the calibration,  estimate $\bmbeta$ by solving 
	\begin{equation} 
		\sum_{i \in \mathcal{S}_1}  \hat{w}_i \bmU_1( \bmbeta; \bmx_{i}, y_i)  = \bmzero. \label{eq: ef beta}
	\end{equation}
\end{enumerate}

%[It might be a good idea to include some explanation why the proposed method is equivalent to the CML method of \cite{chatterjee2016}. I am not 100\% sure about the equivalence. ]

If the benchmark $\bmalpha^*$ is not available from the finite population but can be estimated from an independent external sample, we can use the information from both the original internal sample and the external sample to obtain the benchmark estimate. In practical situations, we may not have access to the raw data of the external sample but often be able to have its summary statistics. 
Suppose that the external {sample} provides a point estimator $\hat{\bmalpha}_2$ and its variance estimator $\bmV_2= \hat{V} ( \hat{\bmalpha}_2)$ for the working reduced model in \eqref{eq-2}.
Then, an estimator of the benchmark $\bmalpha^*$ can be obtained by 
\begin{equation} 
	\hat{\bmalpha}^* =  (\bmV_1^{-1} + \bmV_2^{-1})^{-1}( \bmV_1^{-1}  \hat{\bmalpha}_1 + \bmV_2^{-1} \hat{\bmalpha}_2) 
	\label{gls} 
\end{equation} 
where $\hat{\bmalpha}_1$ and $\bmV_1$ are estimated with the internal  sample $\mathcal{S}_1$. 
Once $\hat{\bmalpha}^*$ is obtained by \eqref{gls}, it replaces $\bmalpha^*$ in the calibration equation in  \eqref{eq-4}.

% \begin{comment} 
% {\bf Comment: Calibration weighting for general parameter estimation } 

% Let $\theta$ be the solution to $E\{ U( \theta; x, y) \} =0$. Let $U ( \alpha ; x) $ be the ``working'' conditional expectation of $U( \theta; x, y) $ given $x$, which is observed in $\mathcal{S}_1$.   Then, (1) we use $\mathcal{S}_1$ to estimate $\alpha$ and then  (2) use above the  empirical likelihood technique to obtain $\hat{w}_i$ incorporating $\hat{\alpha}$ obtained from Step 1. 

% {\color{blue} [ Section ``Bayesian approach'' is now moved to Section \ref{sec:Bayesian} ] }

% \end{comment} 
% ----------------------------

	Similarly to \cite{wu2001model}, the proposed method does not require a ``true'' working model as explained below.
	Let 
	$ \hat{\bmU}_{\text{ext}}( \bmalpha)=0
	$ 
	%\zl{(Hi Hang, I am not sure whether we should use ``$\hat{\bmalpha}_2$'' or not.)}
	be the estimating equation for obtaining $\bmalpha^*$ computed from the external sample $\mathcal{S}_2$. 
	Now, the final estimating function for $\bmbeta$ using the model calibration 
	$ \hat{\bmU}_\text{cal} ( \bmbeta) = \sum_{i \in \mathcal{S}_1} \hat{w}_i {\bmU}_1 ( \bmbeta; \bmx_i, y_i) $
	can be approximated by
	\begin{equation} 
		\hat{\bmU}_\text{cal} ( \bmbeta) \doteq  \hat{\bmU}_1( \bmbeta)  + \bmK  \left\{ \hat{\bmU}_{\text{ext}} ( \bmalpha^*) - \hat{\bmU}_2 ( \bmalpha^*) \right\} 
		\label{appr}
	\end{equation} 
%	\textcolor{blue}{(Dear Dr. Kim and Hang, should we use $\bmalpha^*$ in the above equation (\ref{appr})?)}
	for some $\bmK$ where $\hat{\bmU}_1( \bmbeta)$ and $\hat{\bmU}_2 ( \bmalpha) $ are computed by (\ref{eq-1}) and (\ref{eq-2b}), respectively,  from the internal  sample $\mathcal{S}_1$.  The approximation in \eqref{appr} can be easily derived using the asymptotic equivalence of the calibration estimator and the regression estimator.  
	Thus, even if $E\{ \hat{\bmU}_{\text{ext}} (\bmalpha^*) \}$ is not equal to zero, the solution to $\hat{\bmU}_\text{cal}( \bmbeta) =0$ is consistent as   $E\{  \hat{\bmU}_{\text{ext}} ( \bmalpha) - \hat{\bmU}_2 ( \bmalpha)   \} = 0$ by design.  
	
	%Motivated from the above observation, we can consider  
%	\textcolor{blue}{(Dear Dr. Kim and Hang, it might be better to put the following paragraph into a remark.)}
	\begin{remark} 
	Although the working model $E(Y_i \mid\bmx_{i1}) = m_2(\bmx_{i1};\bmalpha)$ does not need to be correctly specified, we can systematically find $\bmU_2( \bmalpha; \bmx_{i1}, y_i)$ by casting its construction as a missing covariate problem, relying on the regression calibration technique.
	For example, suppose that $\bmx_i = (x_{i1}, x_{i2})$, we set a predictor $\hat{x}_{i2}= {\beta}_0 + {\beta}_1 x_{i1}$, and an estimating equation is written by 
	\begin{equation} 
		\bmU_1 ( \bmbeta; x_{i1}, \hat{x}_{i2}, y_i ) =  \left\{ y_i - m_1 ( x_{i1}, \hat{x}_{i2}; \bmbeta) \right\} \bmh_1 ( x_{i1}, \hat{x}_{i2}; \bmbeta) 
		\label{control} 
	\end{equation} 
	for the control function of the model calibration method where $\bmbeta = (\beta_0,\beta_1)$.  We can either estimate $\bmbeta$ from sample $\mathcal{S}_1$ or use any fixed parameter value as long as the solution to 
	$\sum_{i \in \mathcal{S}_1} d_i \bmU_1 ( \bmbeta; x_{i1}, \hat{x}_{i2}, y_i ) 
	= \bmzero$ is unique. A benchmark estimator of $\bmbeta$ can be obtained using external samples to apply the proposed model calibration method. 
	If we use the control function in \eqref{control}, then we are essentially treating a regression of $y$ on $x_1$ and $\hat{x}_2$ as the ``working'' model for model calibration. 
	This is feasible only when we have direct access to an external sample $\mathcal{S}_2$ in addition to the internal sample $\mathcal{S}_1$. 
\end{remark}

\section{Theoretical properties}\label{sec: theory sec}

In this section, we investigate the asymptotic properties of the the proposed estimator $\hat{\bmbeta}$ to \eqref{eq: ef beta}. Since the population parameters including $\bmbeta_0$ and $\bmalpha^*$ are determined by the finite population of size $N$, we explicitly  use subscript $N$ for those in this section, e.g., $\bmbeta_{0N}$ and $\bmalpha_N^*$, but we omit this subscript for $(d_i,\bmx_i,y_i)$ for simplicity. We consider two scenarios: when $\bmalpha_N^*$ is available from the finite population and when we only have an external sample  to estimate $\bmalpha_N^*$ by the generalized least square in  \eqref{gls}.

% {\color{red}
% [ REFER TO SUPPLEMENTARY MATERIAL ] \\
% Section \ref{sec:MultipleDataIntegration} \\
% Table \ref{t11-5} \\
% Equations
% \eqref{11-eq-1} \eqref{11-eq-2} \eqref{11-eq-3} \eqref{11-eq-4} \eqref{11-eq-5}
% }

\subsection{$\bmalpha_N^*$ is available}

Let $\tilde{d}_i = \hat{N}^{-1}d_i$ where $\hat{N} = \sum_{i\in \mathcal{S}_1}d_i$ is the Horvitz–Thompson estimator of the population size $N$. 
Replacing $d_i$ by $\tilde{d}_i$  in \eqref{eq: Q(d,w)}, we consider the Lagrangian problem that maximizes
$$l(\bmw,\bmlambda,\phi) = \sum_{i\in \mathcal{S}_1}\tilde{d}_i\log(w_i) + \bmlambda^\top \sum_{i\in \mathcal{S}_1}w_i \bmU_2(\bmalpha_N^*;\bmx_{i1},y_i) + \phi\left(\sum_{i\in \mathcal{S}_1}w_i-1\right)$$
where $\bmlambda$ and $\phi$ are the Lagrange multipliers. 
%\textcolor{red}{(Superscript of a superscript should be avoided. Maybe use $\bmU_{1i}^{*}$?)} \zl{(Yes, we have used different notations for $\bmbeta$ and $\bmalpha$. Besides, the abbreviation is avoided.)}

By setting $\partial l(\bmw,\bmlambda,\phi) /\partial\bmlambda = \bmzero$, $\partial l(\bmw,\bmlambda,\phi) /\partial\phi = 0$ and $\partial l(\bmw,\bmlambda,\phi) /\partial w_i = 0$ for  $i\in \mathcal{S}_1$,  we get $\hat{\phi} = -1$ and $\hat{w}_i =\tilde{d}_i\{1-\bmlambda^\top \bmU_2(\bmalpha_N^*;\bmx_{i1},y_i)\}^{-1}$.
Then, the proposed method is equivalent to solving $g(\bmbeta,\bmlambda)=\bmzero$ where 
% \begin{equation}
% g(\bmbeta,\bmlambda) = 
%     \begin{pmatrix}
%     \displaystyle\sum_{i\in \mathcal{S}}\frac{\tilde{d}_i}{1-\bmlambda^\top \bmU_i^{\alpha^*}} \bmU_i^{\alpha^*}\\
%     \displaystyle\sum_{i\in \mathcal{S}}\frac{\tilde{d}_i}{1-\bmlambda^\top \bmU_i^{\alpha^*}} \bmU_i^\beta
%     \end{pmatrix} \label{eq: g eta}
% \end{equation}
\begin{equation}
	g(\bmbeta,\bmlambda) = 
	\begin{pmatrix}
		
		\displaystyle\sum_{i\in \mathcal{S}_1}\frac{\tilde{d}_i}{1-\bmlambda^\top \bmU_2(\bmalpha_N^*;\bmx_{i1},y_i) } \bmU_1(\bmbeta;\bmx_i,y_i)\\    
		\displaystyle\sum_{i\in \mathcal{S}_1}\frac{\tilde{d}_i}{1-\bmlambda^\top \bmU_2(\bmalpha_N^*;\bmx_{i1},y_i) }\bmU_2(\bmalpha_N^*;\bmx_{i1},y_i) 
	\end{pmatrix}. \label{eq: g eta}
\end{equation}
%\textcolor{blue}{(Actually, to get equivalency, we also need $\sum_{i\in \mathcal{S}}\tilde{d}_i(1-\hat{\bmlambda}^\top \bmU_i^{\alpha^*})=1$.) }

Denote the solution to \eqref{eq: g eta} as $\hat{\bmeta} = ( \hat{\bmbeta}{}^{\top},\hat{\bmlambda}{}^{\top} ){}^{\top}$. To investigate asymptotic properties of $\hat{\bmeta}$, we propose the following regularity conditions.
\begin{enumerate}[1.]
	\renewcommand{\labelenumi}{C\arabic{enumi}.}
	\item There exists a compact set $\mathcal{A}$ such that 
	$Z_S=\sup_{\boldsymbol{\alpha}\in\mathcal{A}}\max_{i\in \mathcal{S}_1}\lVert \bmU_2(\bmalpha;\bmx_{i1},y_{i}) \rVert = o_p(n^{1/2})$ and $\bmalpha_N^*\in\mathcal{A}$ for $N\in\mathbb{N}$ where $\lVert\cdot\rVert$ denotes the Euclidean norm and the stochastic order is with respect to the sampling design.  \label{cond: max U1i}
	\item The sampling design satisfies the following convergence results.
	% and the stochastic order is with respect to the sampling design.
	
	\begin{enumerate}[a.]
		
		% \removed{\item $N^{-1}\sum_{i\in \mathcal{S}}{d}_i=1+O_p(n^{-1/2})$.\label{cond: cond2 hat N} \textcolor{blue}{(Since C3 has been replaced, this condition is kind of redundant. Please check.)}}
		\item There exist a compact set $\Omega$ such that $\bmbeta_{0N}\in\Omega$ for $N\in\mathbb{N}$ and an interior point of $\Omega$, $\bmbeta_p$, such that $\lim_{N\to\infty}\bmbeta_{0N}=\bmbeta_p$. \label{cond: beta region}
		\item  There exists a continuous function $\bmU_0(\bmbeta)$ over $\Omega$ such that $\sup_{\boldsymbol{\beta}\in\Omega}\lVert\sum_{i\in \mathcal{S}_1}\tilde{d}_i \bmU_1(\bmbeta;\bmx_i,y_i)-\bmU_0(\bmbeta)\rVert \to 0$  in probability where  $\bmbeta_p$ is the unique solution to $\bmU_0(\bmbeta)=\bmzero$.\label{cond: unique beta}\label{cond: uniform abs}
		\item $\sum_{i\in \mathcal{S}_1}\tilde{d}_i\partial \bmU_1(\bmbeta_{0N};\bmx_{i},y_i)/\partial\bmbeta^{\top} = \bmcalI_{11}+o_p(1)$ where $\bmcalI_{11}$ is non-stochastic and invertible.\label{cond: I 22}
		
		\item $\sum_{i\in \mathcal{S}_1}\tilde{d}_i \bmU_1(\bmbeta_{0N};\bmx_{i},y_i)\bmU_2(\bmalpha_N^*;\bmx_{i1},y_i){}^\top =\bmcalI_{12} + o_p(1)$ where $\bmcalI_{12}$ is non-stochastic.\label{cond: I12}
		\item $\sum_{i\in \mathcal{S}_1}\tilde{d}_i\bmU_2(\bmalpha_N^*;\bmx_{i1},y_i)^{\otimes2}=\bmcalI_{22}+o_p(1)$ where $\bmA^{\otimes2} = \bmA \bmA^\top $ for any matrix $\bmA$ and $\bmcalI_{22}$ is non-stochastic and  positively definitive. \label{cond: variance conv}
	\end{enumerate}\label{cond: I convergence}
	
	\item The sampling design satisfies 
	$$
	n^{1/2}\sum_{i\in \mathcal{S}_1}\tilde{d}_{i}
	\begin{pmatrix}
		\bmU_1(\bmbeta_{0N};\bmx_i,y_i)\\
		\bmU_2(\bmalpha_N^*;\bmx_{i1},y_i)
	\end{pmatrix}\to\mathcal{N}(\bmzero,\bmSigma_{u})
	$$
	in distribution where $\mathcal{N}(\bmzero,\bmSigma_u)$ is a normal distribution with mean zero and covariance matrix $$\bmSigma_u = \begin{pmatrix}
		\bmSigma_{11}&\bmSigma_{12}\\
		\bmSigma_{21}&\bmSigma_{22}
	\end{pmatrix}.$$ \label{cond: CLT}

\end{enumerate}
C\ref{cond: max U1i} is a technical condition to obtain the asymptotic order of $\hat{\bmlambda}$, and a similar condition is also assumed by  \citet{wu2006}; see their condition~C1 for details.  C\ref{cond: I convergence} assumes several convergence results for the two estimating functions. 
% \removed{C\ref{cond: cond2 hat N} shows that $\hat{N}N^{-1}\to1$ in probability, and it is used to get the central limit theorem for $\hat{\bmeta}$. Condition} 
Specifically, C\ref{cond: beta region} shows the parameter space of the finite population parameter $\bmbeta_{0N}$, and the convergence of $\bmbeta_{0N}$  can be satisfied under regularity conditions. Condition~C\ref{cond: unique beta} is necessary to show $\hat{\bmbeta}-\bmbeta_p\to0$ in probability, then  $\hat{\bmbeta}-\bmbeta_{0N}\to0$ in probability, coupled with C\ref{cond: beta region}. 
Conditions C\ref{cond: I 22}--C\ref{cond: variance conv} guarantee the central limit theorem for $\hat{\bmeta}$. 
Note that $\bmcalI_{22}$ is symmetric by C\ref{cond: variance conv}, but $\bmcalI_{11}$ in C\ref{cond: I 22} may be asymmetric for a certain estimating function $\bmU_1(\bmbeta;\bmx,y)$. 
Condition C\ref{cond: CLT} is satisfied under regularity conditions for general sampling designs; see \citet[Section~1.3]{fuller2009} for details.

% \begin{lemma}\label{theo: lambda cons}
% 	Suppose that conditions C\ref{cond: max U1i}\removed{, C\ref{cond: cond2 hat N}}, C\ref{cond: variance conv} and C\ref{cond: CLT} hold. Then, $\lVert\hat{\bmlambda}\rVert = O_p(n^{-1/2})$.
% \end{lemma}
% The proof of Lemma~\ref{theo: lambda cons} is presented in \ref{proof: lambda constant} of the Supplementary Material. Lemma~\ref{theo: lambda cons} establishes that $\hat{\bmlambda} = o_p(1)$, and it is essential to investigate the asymptotic distribution of $\hat{\bmeta}$.

% \begin{lemma}\label{theo: beta cons}
% 	Suppose that conditions C\ref{cond: max U1i}, \removed{C\ref{cond: cond2 hat N}}C\ref{cond: beta region}--C\ref{cond: variance conv} and C\ref{cond: CLT}  hold. Then, $\hat{\bmbeta}-\bmbeta_{0N}=o_p(1)$.
% \end{lemma}
% The proof of Lemma~\ref{theo: beta cons} is presented in \ref{proof: beta constant} of the Supplementary Material. By Lemmas~\ref{theo: lambda cons}--\ref{theo: beta cons}, we conclude that $\hat{\bmeta}-\bmeta_0=o_p(1)$ where $\bmeta_0 = (\bmzero{}^\top,\bmbeta_{0N}{}^\top){}^\top$. Thus, we can use Taylor expansion to establish the following central limit theorem for $\hat{\bmeta}$.

\begin{theorem}\label{theo: CLT}
	Suppose that conditions C\ref{cond: max U1i}--C\ref{cond: CLT} hold. Then,  
	% \begin{equation}
	$
	n^{1/2}(\hat{\bmeta} - \bmeta_0)\to \mathcal{N}(\bmzero,\bmSigma_\eta)\notag
	$
	% \end{equation}
	in distribution where $
	\bmSigma_\eta = \bmcalI^{-1}\bmSigma_u(\bmcalI^{-1})^\top 
	$ and 
	$$
	\bmcalI = \begin{pmatrix}
		\bmcalI_{11}&\bmcalI_{12} \\
		\bmzero&\bmcalI_{22}
	\end{pmatrix}.
	$$
\end{theorem}
The proof of Theorem~\ref{theo: CLT} is presented in Appendix~\ref{proof: CLT}.
By Theorem~\ref{theo: CLT}, we can obtain that 
$
n^{1/2}(\hat{\bmbeta}-\bmbeta_{0N})\to\mathcal{N}(\bmzero,\bmSigma_\beta)
$
% \label{result5}
in distribution where 
% \begin{equation}\label{eq: var beta theo 1}
$$
\bmSigma_\beta = \bmcalI_{11}^{-1}\bmSigma_{11}(\bmcalI_{11}^{-1})^{\T} -  \bmcalI_{11}^{-1}\bmcalI_{12}\bmcalI_{22}^{-1} \bmSigma_{21}(\bmcalI_{11}^{-1})^{\T} - \bmcalI_{11}^{-1}\bmSigma_{12}\bmcalI_{22}^{-1} \bmcalI_{12}^{\T}(\bmcalI_{11}^{-1})^{\T}
+\bmcalI_{11}^{-1}\bmcalI_{12}\bmcalI_{22}^{-1} \bmSigma_{22}\bmcalI_{22}^{-1} \bmcalI_{12}^{\T}(\bmcalI_{11}^{-1})^{\T}
$$
% \end{equation} 
and $\bmSigma_{11}$ and $\bmSigma_{22}$ correspond to the asymptotic variances of $n^{1/2}\sum_{i\in \mathcal{S}_1}\tilde{d}_{i}\bmU_1(\bmbeta_{0N};\bmx_{i},y_i)$ and $n^{1/2}\sum_{i\in \mathcal{S}_1}\tilde{d}_{i}\bmU_2(\bmalpha_N^*;\bmx_{i1},y_i)$, respectively. {Furthermore, we have the following result regarding the optimality of $\bmU_2(\bmalpha_N^*;\bmx_{i1},y_i)$.}
\begin{corollary}\label{cor: optimal var}
	Suppose that the conditions in Theorem~\ref{theo: CLT} hold. For a fixed estimating function $\bmU_1(\bmbeta;\bmx,y)$,   $\hat{\bmbeta}$ is optimal if  $\bmcalI_{12}\bmcalI_{22}^{-1}\bmU_2(\bmalpha_N^*;\bmx_{1},y) = E\{\bmU_1(\bmbeta_{0N};\bmx,y)\mid \bmx_{1},y\}$ holds almost surely for the working reduced model, where $\bmx=(\bmx_1,\bmx_2)$, and the expectation is taken with respect to the super-population model.
\end{corollary}
{The proof of Corollary~\ref{cor: optimal var} is relegated to Appendix~\ref{sec app proof cor opti}. Corollary~\ref{cor: optimal var} presents a sufficient condition on the reduced model to guarantee an optimal estimator $\hat{\bmbeta}$ if the working model is correctly specified. That is, even if we do not require that the reduced model is correctly specified for consistency, the efficiency gain is guaranteed only under the correct model specification. By Corollary~\ref{cor: optimal var}, an optimal estimator of $\bmalpha_N^*$ can be obtained by solving $E\{\bmU_1(\bmbeta_{0N};\bmx,y)\mid \bmx_{1},y\}=\bmzero$.}

Under regularity conditions, it can be shown that  $\bmSigma_\beta = \bmcalI_{11}^{-1}(\bmSigma_{11} - \bmSigma_{12}\bmSigma_{22}^{-1} \bmSigma_{21})(\bmcalI_{11}^{-1})^\top $ for simple random sampling with or without replacement. Since $\bmcalI_{11}^{-1}\bmSigma_{11}(\bmcalI_{11}^{-1})^\top $ is the asymptotic variance of $n^{1/2}(\hat{\bmbeta}_m-\bmbeta_{0N})$ where $\hat{\bmbeta}_m$ solves $\sum_{i\in \mathcal{S}_1}d_i\bmU_1(\bmbeta;\bmx_i,y_i)=0$, the proposed  approach achieves efficient estimation under simple random sampling; see  \ref{SS: hat beta variance} of the Supplementary Material for details. 
%\textcolor{red}{(One remaining question is whether our proposed method is asymptotically equivalent to the method of \cite{chatterjee2016} under SRS. If they are different, which one is better? -JK) } 

%\textcolor{red}{Also, the uncertainty in $\hat{\bmalpha}^*$ is ignored in the proof. We also need to prove a theory for the case when $\hat{\alpha}^*$ is estimated by \eqref{gls}). We might want to use bootstrap approach in this case. See also the Bayesian updating method of \cite{lyddon2019}.}

%\textcolor{blue}{Yes, the above result does not consider the uncertainty of $\alpha^*$, and the corresponding theoretical results are provide in the next section.}

\subsection{An external estimator $\hat{\bmalpha}_2$ is available}

When $\bmalpha^*$ is not available but an external sample is available to get $\hat{\bmalpha}^* $ in \eqref{gls}, we consider
\begin{equation}
	\tilde{g}(\bmeta) = 
	\begin{pmatrix}
		\displaystyle\sum_{i\in \mathcal{S}_1}\frac{\tilde{d}_i}{1-\bmlambda^\top \bmU_2(\hat{\bmalpha}^*;\bmx_{i1}, y_i)} \bmU_1(\bmbeta;\bmx_i,y_i)\\
		\displaystyle\sum_{i\in \mathcal{S}_1}\frac{\tilde{d}_i}{1-\bmlambda^\top \bmU_2(\hat{\bmalpha}^*;\bmx_{i1}, y_i)} \bmU_2(\hat{\bmalpha}^*;\bmx_{i1}, y_i)
		
	\end{pmatrix}. \label{eq: tilde g eta}
\end{equation}
Denote $\tilde{\bmeta}$ to be the solution of  $\tilde{g}(\bmeta) = \bmzero$. Then, the following additional assumptions are required to get the asymptotic properties for $\tilde{\bmeta}$.
\begin{enumerate}[1.]
	\renewcommand{\labelenumi}{C\arabic{enumi}.}
	\setcounter{enumi}{3}
	\item  $\sum_{i\in \mathcal{S}_1}\tilde{d}_i\partial \bmU_2(\bmalpha;\bmx_{i1},y_i)/\partial\bmalpha^\top =\bmcalI(\bmalpha)+o_p(1)$ uniformly for $\bmalpha\in\mathcal{A}$ where $\bmcalI(\bmalpha)$ is non-stochastic. Besides, there exists an invertible matrix $\bmcalI_0$ such that $\lim_{N\to\infty}\bmcalI(\bmalpha_N^*) = \bmcalI_0$.\label{cond: I alpha}
	\item The sampling design and the external sample satisfy the following convergence results.
	\begin{enumerate}[a.]
		\item Both $\hat{\bmalpha}_1$ and $\hat{\bmalpha}_2$ are consistent for $\bmalpha^*$. \label{eq: cons alphas}
		\item $\bmV_1$ and $\bmV_2$ are design consistent variance estimators of $\hat{\bmalpha}_1$ and $\hat{\bmalpha}_2$, respectively. \label{cond: var consist alphas}
		\item $\bmV_1^{-1}$, $\bmV_2^{-1}$, and $(\bmV_1^{-1} + \bmV_2^{-1})^{-1}$ exist in probability. \label{cond: inverse exis Vs}
		\item $(\bmV_1^{-1} + \bmV_2^{-1})^{-1}\bmV_2^{-1}=\bmW+o_p(1)$ where $\bmW$ is non-stochastic. \label{cond: W2 convergence}
		\item There exists a scaling function $\gamma(n)$ such that $\gamma(n)(\hat{\bmalpha}_2-\bmalpha^*)\to\mathcal{N}(0,\bmSigma_2)$ in distribution where $\bmSigma_2$ satisfies $\gamma(n)^2\bmV_2 = \bmSigma_2 + o_p(1)$.\label{cond: CLT for alpha hat 2}
	\end{enumerate} \label{cond: consistent alphas}
\end{enumerate}
C\ref{cond: I alpha} is used to obtain the asymptotic order and the variance of $\hat{\bmalpha}^* - \bmalpha_N^*$, and a similar condition was used by \citet{yuan1998asymptotics}. C\ref{eq: cons alphas} and C\ref{cond: var consist alphas} assume the consistency of  $\hat{\bmalpha}_2$ and $\bmV_2$ obtained by an external sample. For the consistency of $\hat{\bmalpha}_1$, a sufficient condition is similar with C\ref{cond: unique beta}. The design consistency of the variance estimator $\bmV_1$ can be obtained under general sampling designs; see \citet[Chapter~1]{fuller2009} for details. 
C\ref{cond: inverse exis Vs} guarantees the existence of $\hat{\bmalpha}^*$ for the proposed method. C\ref{cond: CLT for alpha hat 2} shows the central limit theorem with respect to the summary statistic $\hat{\bmalpha}_2$, and it is used to derive a similar result as C\ref{cond: CLT} with $\bmalpha^*$ replaced by $\hat{\bmalpha}^*$. Specifically, the convergence rate of $(\hat{\bmalpha}_2-\bmalpha^*)$ is $\gamma(n)^{-1}$, which is determined by the external sample. 

The following theorem establishes an asymptotic distribution similar to that in C\ref{cond: CLT}.
\begin{theorem}\label{theo: CLT hat alpha *}
	Suppose that conditions  C\ref{cond: max U1i}\removed{, C\ref{cond: cond2 hat N}} and C\ref{cond: CLT}--C\ref{cond: consistent alphas} hold. Then,  
	$$n^{1/2}\sum_{i\in \mathcal{S}_1}\tilde{d}_{i}
	\begin{pmatrix}
		\bmU_1(\bmbeta_0;\bmx_i,y_i) \\
		\bmU_2(\hat{\bmalpha}^*;\bmx_{i1},y_i)
	\end{pmatrix}\to\mathcal{N}(\bmzero,\tilde{\bmSigma}_{u})$$
	in distribution where  
	$$
	\tilde{\bmSigma}_{u} = \begin{pmatrix}
		\tilde{\bmSigma}_{11}&\tilde{\bmSigma}_{12}\\
		\tilde{\bmSigma}_{21}&\tilde{\bmSigma}_{22}
	\end{pmatrix}
	$$  
	Case 1. Specifically, 
	%  \begin{enumerate}
	%   \renewcommand{\labelenumi}{Case \arabic{enumi}}
	%   \item 
	if there exists a non-stochastic matrix $\bmSigma_c$ such that $n\bmV_2=\bmSigma_c+o_p(1)$, then   $\tilde{\bmSigma}_{11} = \bmSigma_{11}$, $\tilde{\bmSigma}_{12}=\bmSigma_{12}(\bmcalI_0^{-1})^{\top}\bmW^{\top}\bmcalI_0^{\top}$, $\tilde{\bmSigma}_{21} = \tilde{\bmSigma}_{12}^{\T}$ and $\tilde{\bmSigma}_{22} = \bmcalI_0\bmW
	\{\bmSigma_{c} +\bmcalI_0^{-1}\bmSigma_{22}(\bmcalI_0^{-1})^{\T}  \}\bmW^{\T}\bmcalI_o^{\T}$; \label{theo: 4.1}
	
	\noindent Case 2. If $ \bmW =0$, then $\tilde{\bmSigma}_{ij} = \bmzero$ for $(i,j)\neq(1,1)$ and $\tilde{\bmSigma}_{11} = \bmSigma_{11}$.
	% \end{enumerate}    
\end{theorem}

The proof of Theorem~\ref{theo: CLT hat alpha *} is presented in Appendix~\ref{SS: PROOF T4}. 
For Case 1, if $\hat{\bmalpha}_2$ estimated from an external sample is much more efficient than $\hat{\bmalpha}$ in the sense of $(\hat{\bmalpha}_2-\bmalpha_N^*) = o_p(n^{-1/2})$, then $ \bmW$ is an identity matrix and $\tilde{\bmSigma}_{ij} = \bmSigma_{ij}$ for $i,j=1,2$. 
Thus, we can ignore the variability of the summary statistic $\hat{\bmalpha}_2$ from the external sample and get the same asymptotic distribution as in C\ref{cond: CLT}. 
Although the asymptotic distributions are the same, C\ref{cond: CLT} with known $\bmalpha_N^*$ is not a special case of Theorem~\ref{theo: CLT hat alpha *} since $\hat{\bmalpha}_2=\bmalpha_N^*$ has zero variance, which violates C\ref{cond: inverse exis Vs}--C\ref{cond: CLT for alpha hat 2}. 
On the other hand, if $(\hat{\bmalpha}_2-\bmalpha_N^*)\asymp n^{-1/2}$ in probability, then $\hat{\bmalpha}_2$ is as efficient as $\hat{\bmalpha}_1$. 
Thus, $ \bmW$ is not an identity matrix nor a zero matrix,  and  the proposed method is more efficient than one replacing $\bmalpha^*$ by $\hat{\bmalpha}^*=\hat{\bmalpha}_2$ due to the extra information provided by the external sample. 
It is trivial that we cannot use $\hat{\bmalpha}_1$ to replace $\bmalpha^*$ in \eqref{eq-4}; otherwise, we get $\hat{w}_i=\tilde{d}_i$, and \eqref{eq: ef beta} is equivalent to the traditional estimation equation $\sum_{i\in \mathcal{S}_1}\tilde{d}_i\bmU_1(\bmbeta;\bmx_i,y_i)=\bmzero$ without calibration. 
If  $\hat{\bmalpha}_2$ is much less efficient than $\hat{\bmalpha}_1$ in terms  of convergence rate, then we should not use such an external sample for the proposed method because  $\hat{\bmalpha}^*-\bmalpha^* = \hat{\bmalpha}_1 - \bmalpha^* + o_p(n^{-1/2})$ and $n^{1/2}\sum_{i\in \mathcal{S}_1}\tilde{d}_i \bmU_2(\hat{\bmalpha}^*;\bmx_{i1},y_i)=o_p(1)$; see \ref{SS: PROOF T4} of the Supplementary Material for details.
% \begin{lemma}
% Under the conditions of Theorem~\ref{theo: CLT hat alpha *}, if $\bmSigma_c$ is not a zero matrix, then $\bmSigma_u - \tilde{\bmSigma}_u$ is positively definitive. 
% \end{lemma}
By  C\ref{cond: consistent alphas}, we can obtain the same consistency results in  Lemmas~\ref{theo: lambda cons}--\ref{theo: beta cons} for \eqref{eq: tilde g eta} under the same conditions. Thus, by Theorem~\ref{theo: CLT hat alpha *}, we obtain the following  asymptotic distribution for $\tilde{\bmeta}$.
\begin{corollary}\label{coro: tilde eta}
	Suppose that conditions C\ref{cond: max U1i}--C\ref{cond: consistent alphas} hold. Then, we have 
	% \begin{equation}
	$
	n^{1/2}(\tilde{\bmeta} - \bmeta_0)\to \mathcal{N}(0,\tilde{\bmSigma}_\eta)\notag
	$
	% \end{equation}
	in distribution where $\tilde{\bmSigma}_\eta = \bmcalI^{-1}\tilde{\bmSigma}_u(\bmcalI^{-1})^\top $, the form of $\bmcalI$ is in Theorem~\ref{theo: CLT}, and the form of $\tilde{\bmSigma}_\eta$ is in Theorem~\ref{theo: CLT hat alpha *}.
\end{corollary}

% {
% The proposed calibration method can also been used to make inference for the parameters in a super-population model; see Section~\ref{sec: super-popu model} for details. 
% }

\begin{remark}
It is worthy pointing out that when deriving the asymptotic properties in this section,  we do not consider the weighting adjustments such as nonresponse adjustment, trimming, and raking. However, those weighting adjustments are commonly used in survey sampling. Thus, it is a promising research topic to generalize the proposed method incorporating those weighting adjustments.
\end{remark}

\section{Multiple data integration} 
\label{sec:MultipleDataIntegration}

We now consider regression analysis combining partial information from  external samples.  To explain the idea, Table \ref{t11-5} shows an example data structure with three data sources ($A$, $B$, $C$) where Sample $A$  contains  all the observations  while samples $B$ and $C$ contain partial observations. 
% We assume that sample A is representative but samples B and C are not necessarily representative of the population. 
%We discuss a measurement error model approach to data integration in \S 2.3.1. 

%\lipsum[2]

\begin{table}[tbh]
	%\label{table:11-5} 
	\caption{\label{t11-5} Data structure for survey integration}
	\begin{center} 
		\begin{tabular}{c|c|cccc}
			\hline 
			Sample  & Sampling Weight      & $z$ &   $x_1$ &  $x_{2}$  & $y$  \\
			\hline 
			$A$     &  $d_{a}$ & \checkmark &  \checkmark	 & \checkmark & \checkmark  \\ %\hline
			$B$    &  $d_{b} $ & \checkmark &  \checkmark      &    & \checkmark    \\ %\hline
			$C$    &  $d_c$ &   \checkmark &     &  \checkmark   & \checkmark    \\ %\hline
			\hline 
		\end{tabular}
	\end{center} 
\end{table}

Under the setup of Table 1, suppose that we are interested in estimating the parameters in the regression model 
$ E( Y | x_1, x_2) = m_1( \beta_0 + \beta_1 x_1 + \beta_2 x_2)$ where $m_1( \cdot)$ is known but $\bmbeta = (\beta_0, \beta_1, \beta_2)$ is unknown. The estimating equation for $\bmbeta$ using sample $A$ can be written as 
\begin{equation} 
	\hat{\bmU}_a(\bmbeta) \equiv \sum_{i \in A} d_{a,i} \{ y_i - m( x_{i1}, x_{i2}; \bmbeta) \} \bmh(x_{i1}, x_{i2}; \bmbeta) = \bmzero, 
	\label{11-eq-1}
\end{equation}
for some $\bmh (x_{i1}, x_{i2}; \bmbeta)$ such that $\hat{\bmU}_a( \bmbeta)$ is linearly independent almost everywhere. 

Now, we wish to incorporate the partial information from sample $B$. To do this, suppose that we have a ``working'' model for $E(Y | x_1, z)$: 
\begin{equation} 
	E( Y | x_1, z) = m_2( x_1, z; \bmalpha) 
	\label{11-eq-2} 
\end{equation} 
for some $\bmalpha$. Note that, since $(z_i, x_{1i}, y_i)$ are observed,  we can use sample $B$ to estimate ${\bmalpha}$ by solving  $\sum_{i \in B} d_{b,i} \bmU_b( \bmalpha; x_{i1}, z_i, y_i)  = \bmzero$ for some $\bmU_b$ satisfying $E\{  \bmU_b( \bmalpha; x_{1},z,  Y)  | x_1, z\} =\bmzero$ under the working model \eqref{11-eq-2}.

Similarly, to incorporate the partial information from sample $C$, suppose that we have a ``working'' model for $E(Y | x_2, z)$: 
\begin{equation} 
	E( Y | x_2, z) = m_3( x_2, z; \bmgamma) 
	\label{11-eq-3}
\end{equation} 
for some $\bmgamma$. We can also  construct an unbiased estimating equation $\sum_{i \in C} d_{c,i}  \bmU_c( \bmgamma; x_{i2},z_i,  y_i)  = \bmzero$ for some $\bmU_c$ satisfying $E\{  \bmU_c( \bmgamma; x_{2},z,  Y)  \mid x_2, z\} =\bmzero$ 
under the working model \eqref{11-eq-3}.  
Once $\hat{\bmalpha}$ and $\hat{\bmgamma}$ are obtained, we can use this extra information to improve the efficiency of $\hat{\bmbeta}$ in \eqref{11-eq-1}. To incorporate the extra information, we can formulate it as maximizing 
$Q(\bmd_a, \bmw)= \sum_{i \in A}d_{a,i} \log \left(w_i\right) $ 
subject to $\sum_{i \in A} w_i=N$ and 
\begin{equation} 
	\sum_{i \in A}  w_i \left[  \bmU_b( \hat{\bmalpha}; x_{i1},z_i,  y_i), \bmU_c( \hat{\bmgamma};x_{i2}, z_i, y_i)\right]   = \bmzero
	\label{11-eq-4}
\end{equation}
where $\bmd_a$ and $\bmw$ are sets containing the sampling weights and calibration weights with respect to sample $A$.
Constraint \eqref{11-eq-4} incorporates the extra information. 
Once the solution $\hat{w}_i$ is obtained, we can use 
% \begin{equation} 
$ \sum_{i \in A}  \hat{w}_i  \{ y_i - m( x_{i1}, x_{i2}; \bmbeta) \} \bmh(x_{i1}, x_{i2}; \bmbeta) = \bmzero $
% \label{11-eq-5}\notag
% \end{equation}
to estimate $\bmbeta$. The asymptotic results can be obtained similarly in Section \ref{sec: theory sec}. 

\begin{remark}
In this paper, we implicitly assume that the populations for the internal sample and the external samples are the same, but it is possible that those populations differ in some scenarios. For example, the external estimator $\hat{\bmalpha}$ may be obtained based on  a non-probability sample, whose sampling frame differs from the one for the probability sample due to the coverage bias in many opt-in surveys. There are several data integration methods incorporating information from heterogeneous populations. For example, \citet{Tayloretal2022} proposed to use ratios of coefficients to  incorporate the external information under regularity conditions even when the populations for the internal and external samples differ. See also \citet{ZhaiHan2022} and \citet{sheng2022synthesizing} for penalized approaches when incorporating external information from heterogeneous populations. The aforementioned existing methods do not take the complex sampling properties into consideration, so it is promising to investigate data integration for heterogeneous populations under survey sampling in a future project. 
\end{remark}
% %%%%%%%%%%%%%%%%%%%%%%%%%%%%%%%%%%%%%%%%%%%
% \newpage 

	\section{Simulation study}
\label{sec:simulation_study}

To evaluate the finite sample  performance of the proposed estimator, we conducted   simulation studies assuming several scenarios.
We generated a finite population of size $N=100{,}000$, each record consisting of auxiliary variables $\bmx_i=(x_{i1},x_{i2})^\top$ of length $p=2$ and a response variable $y_i$. We assume  that  $(\bmx_i,y_i)$ is available for the internal sample $\mathcal{S}_1$ while  only $(x_{i1},y_i)$ is available for the external sample $\mathcal{S}_2$. %For comparison, we assume a linear regression model and a logistic regression model for $\bmU_1(\bmbeta; \bmx_i, y_i)$ and consider scenarios where $x_{i1}$ and $x_{i2}$ are either independent or dependent. 

%\subsection{Linear regression model setup}\label{subsec: linear}

We  evaluate the performance of the  proposed estimator under  a linear regression  setup. In this case,  we are interested in making statistical inference for $\bmbeta = (\beta_0,\beta_1,\beta_2)^\top$ that solves 
% \begin{equation}
$\sum_{i=1}^N (y_i-\beta_0 - \beta_1  x_{i1}- \beta_2 x_{i2})(1, x_{i1}, x_{i2})^\top = \bmzero$.

% \end{equation}

First, we consider two scenarios to generate covariates for the finite population: (i) $x_{i1}\sim N(3,1)$ and $x_{i2} \sim N(11,6.5^2)$ where $x_{i1}$ and $x_{i2}$ are independent;  (ii) $x_{i1} \sim N(3,1)$ and $x_{i2} = x_{i1}^2 +\epsilon_i$ with $\epsilon_i\sim N(0,1)$. The simulation parameters are  chosen such that the marginal mean and variance of $x_{i2}$ are similar in the independent and the dependent settings. Second, the response variable is generated as 
$Y_i = \mu_i + \varepsilon_i$ with $\mu_i=1 + 2 x_{i1} + x_{i2}$ under two scenarios: (i) homogeneous variance with $\varepsilon_i\sim N(0,9)$ and (ii) heterogeneous variance with $\varepsilon_i\mid \bmx_i \sim N(0,\sigma_i^2)$ with $\sigma_i=0.2\lvert\mu_i\rvert$.
Third, we consider two sampling designs to generate a probability sample $\mathcal{S}_1$ of (expected) size $n_1 = 1{,}000$: (i) simple random sampling  without replacement (SRS), and (ii) Poisson sampling with inclusion probabilities satisfying $\pi_{1i}\propto(y_i-\min\{ y_i:i=1,\ldots,N\}+10)^{1/2}$ and $\sum_{i=1}^N\pi_{1i}=n_1$. Last, we consider two sampling designs to generate an external sample $\mathcal{S}_2$ of (expected) size $n_2 = 10{,}000$: (i) SRS and (ii) Poisson sampling with inclusion probabilities satisfying $\pi_{2i}\propto\{1 + \exp(0.2x_{i1} + 0.1x_{i2}-0.6)\}^{-1}$ and $\sum_{i=1}^N\pi_{2i}=n_2$. It is worthy pointing out that the sampling design for the internal sample is informative \citep{pfeffermann1993role} under Poisson sampling, so ignoring the design feature may result in erroneous inference. 

For the proposed estimator, we consider a working reduced model, 
% \begin{equation}
$\sum_{i\in\mathcal{S}_2}\pi_{i2}^{-1}(y_i-\alpha_0-\alpha_1 x_{i1})(1, x_{i1})^\top = \bmzero$, 
% \label{eq: working}\notag
% \end{equation}
whose solution is denoted as $\hat{\bmalpha}_2$. Based on the external sample $\mathcal{S}_2$, we assume that a point estimator $\hat{\bmalpha}_2$ and its variance estimator $\bmV_2= \hat{V} ( \hat{\bmalpha}_2)$ are available as discussed in Section~\ref{sec:Proposed approach}. Linearization is adopted to obtain a variance estimator $\bmV_2$; see the proof of Theorem~\ref{theo: CLT} in  \ref{proof: CLT} of the Supplementary Material for details. 

In the simulation study, the proposed estimator is  compared with the constrained maximum likelihood (CML) estimator \citep{chatterjee2016}. We assume a normal distribution for the likelihood function, i.e., $y_i \mid \bmx_i \sim N\{ (1, \bmx_i^\top) \bmbeta, \sigma_\text{full}^2 \}$. 
We also suppose that an analyst assumes $y_i \mid \bmx_{i1} \sim N\{ (1,  x_{i1}) \bmalpha, \sigma_\text{red}^2 \}$ for the working reduced model.  
% and the ``reduced model''  is also assumed to be normal as $g(y_i\mid  \bmxoi;\bmtheta_r) = (2\pi\sigma_\text{red}^2)^{-1/2}\exp\{-(y_i- \bmxoi^\top \bmalpha)^2/(2\sigma_\text{red}^2)\}$ where $\bmtheta_f = (\bmbeta^\top ,\sigma_\text{full}^2)^\top $, $\bmxoi = (1, x_{i1})^\top $, and $\bmtheta_r = (\bmalpha^\top ,\sigma_\text{red}^2)^\top $. Furthermore, we assume that the solution to
% \begin{equation}
% % \sum_{i=1}^N\begin{pmatrix}
% % \dfrac{y_i-\bmxoi^\top \bmalpha}{\sigma_\text{red}^2}\bmxoi\\[0.5cm]
% % -\dfrac{1}{2\sigma_\text{red}^2} + \dfrac{(y_i-\bmxoi^\top \bmalpha)^2}{2(\sigma_\text{red}^2)^2}
% % \end{pmatrix}=0
% \sum_{i=1}^N \dfrac{y_i-(1,x_{i1}) \bmalpha}{\sigma_\text{red}^2} \begin{pmatrix}
% 1 \\ x_{i1}
% \end{pmatrix}
%  = \bmzero \BB \text{ and } \BB 
% \sum_{i=1}^N -\dfrac{1}{2\sigma_\text{red}^2} + \dfrac{\{ y_i-(1,x_{i1}) \bmalpha\}^2}{2 \sigma_\text{red}^4} = \bmzero. \notag
% \end{equation} 
% is known, and denoted it to be $\bmtheta_r^*$. 
% This is the original Chatterjee's estimator \citep{chatterjee2016}; 
See \ref{ss: linear CML} of the Supplementary Material for the computation details. 
% The full model is correctly specified when the random errors following a normal distribution, but they are not when we use a $t$-distribution instead.
We consider the CML estimator under the setting where the extra information of $(y_i,x_{i1})$ is available for an external sample, not for the entire population. 

% We use the same full and reduced models in Ch1, but $\bmtheta_r^*$ is not available. Instead, we assume that an external sample $S_e$ is available. Then, we can   obtain $\hat{\bmtheta}_r^*$ by combining the two sources together in a similar way as \eqref{gls}).

\begin{figure}[!ht]
	\centering %This figure is generated by first_simulation_strict_chatt_plot_v1.R in the server
	\includegraphics[width=\textwidth]{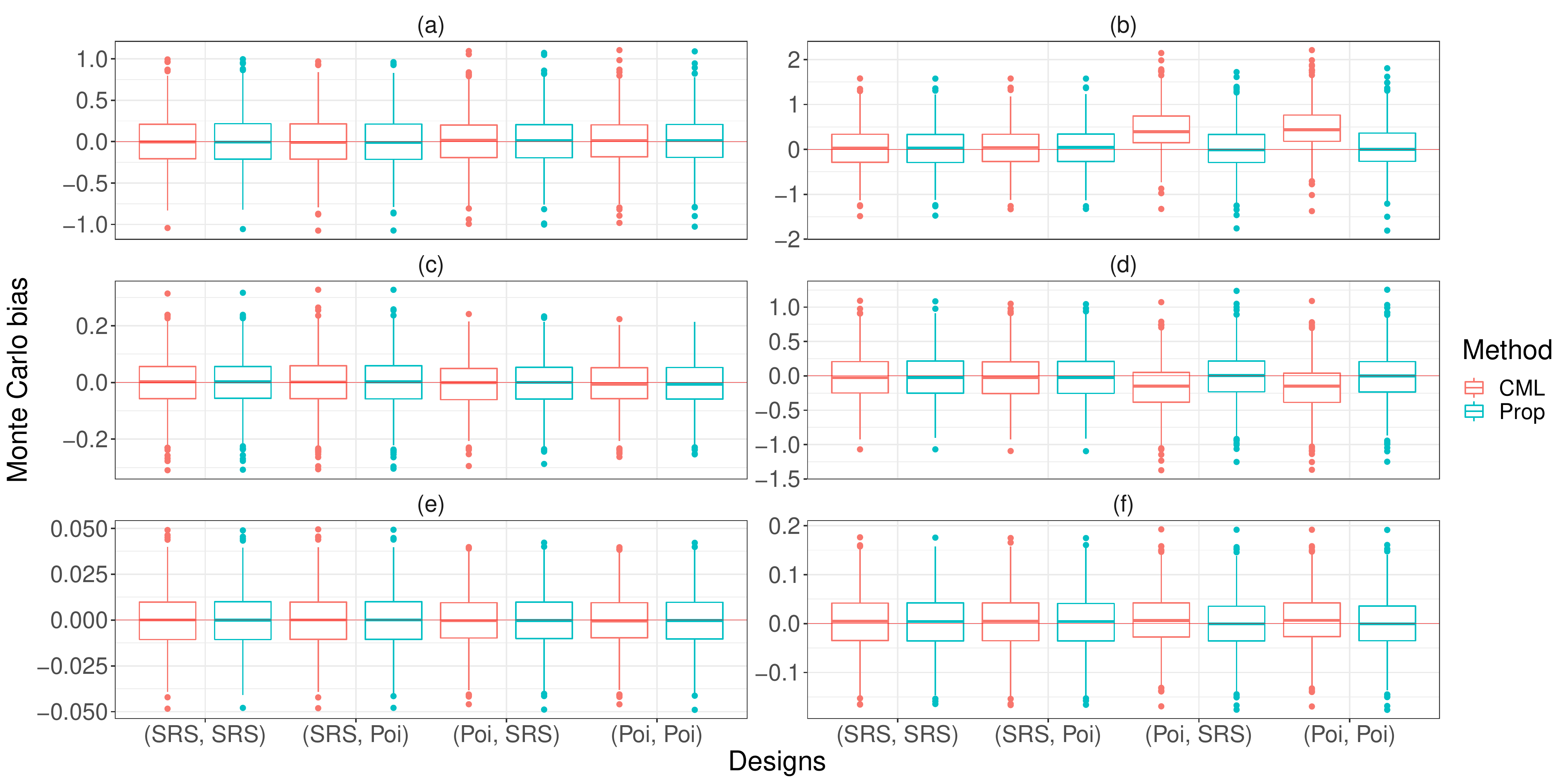}
	\caption{Monte Carlo bias of   the proposed and CML estimators  based on 1,000 Monte Carlo simulations under the homogeneous variance setup. The first to the third rows stand for the Monte Carlo bias for estimating $\beta_0$, $\beta_1$ and $\beta_2$, respectively. The three plots, including (a), (c) and (e), in the left column show the results when the auxiliary variables are independently generated, and those, including (b), (d) and (f), in the right column are for the case when the auxiliaries are dependent.  ``CML'' and ``Prop'' stands for the CML estimator and the proposed estimator, respectively. The first design in the parenthesis is used to generate the internal sample $\mathcal{S}_1$, and the second one to generate the external sample $\mathcal{S}_2$. ``Poi'' represents Poisson sampling.}
	\label{fig: SRS}
\end{figure}

\begin{figure}[!ht]
	\centering %This figure is generated by first_simulation_strict_chatt_plot_v1.R in the server
	\includegraphics[width=\textwidth]{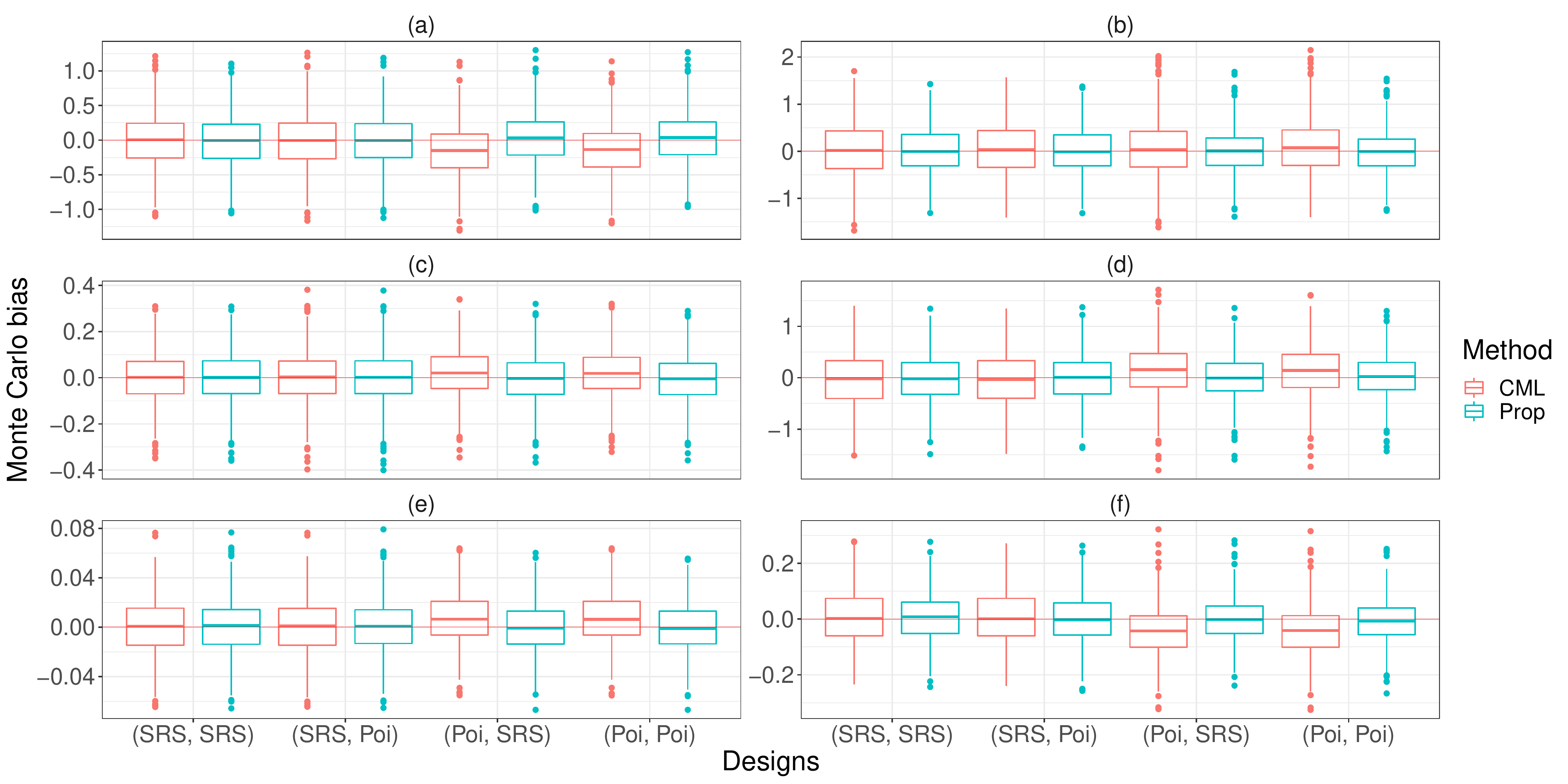}
	\caption{Monte Carlo bias of   the proposed and CML estimators  based on 1,000 Monte Carlo simulations under the heterogeneous variance setup. The first to the third rows stand for the Monte Carlo bias for estimating $\beta_0$, $\beta_1$ and $\beta_2$, respectively. The three plots, including (a), (c) and (e), in the left column show the results when the auxiliary variables are independently generated, and those, including (b), (d) and (f), in the right column are for the case when the auxiliaries are dependent.  ``CML'' and ``Prop'' stands for the CML estimator and the proposed estimator, respectively. The first design in the parenthesis is used to generate the internal sample $\mathcal{S}_1$, and the second one to generate the external sample $\mathcal{S}_2$. ``Poi'' represents Poisson sampling.}
	\label{fig: POI}
\end{figure}

We conduct $M=1{,}000$ Monte Carlo simulations, and Figures~\ref{fig: SRS} and \ref{fig: POI} show the  Monte Carlo bias of  the proposed and CML estimators for the homogeneous and heterogeneous variance setups, respectively. From Figure~\ref{fig: SRS}, when the variance of the error term is homogeneous and the internal sample is generated by SRS, the proposed estimator performs approximately the same as CML estimator in terms of Monte Carlo bias and variance. However, when the auxiliaries are correlated and the internal sample is generated by Poisson sampling, the CML estimator is questionable, since its model is wrongly specified under the informative Poisson sampling design. For example, the Monte Carlo bias of the CML estimator is not negligible when estimating $\beta_0$ and $\beta_1$. Because the proposed estimator incorporates the design features, its performance is satisfactory for all setups. As shown in Figure~\ref{fig: POI}, even when the internal sample is generated by SRS,  the CML estimator is slightly less efficient than the proposed estimator. The reason is that the CML estimator fails to take the heterogeneous variance into consideration, but the proposed estimator does not make any distribution assumption. When the internal sample is generated by an informative Poisson sampling design, the CML performs poorly,  since it is not unbiased, and since its variance is larger than the proposed estimator. 

Table~\ref{tab: SRS CI} shows the coverage rate of a 95\% confidence interval for the proposed estimator under different settings.  \citet{chatterjee2016} only investigated the theoretical properties of their estimator when the population-level information is available. Thus, no interval estimator can be provided if only an external sample is available. By Table~\ref{tab: SRS CI}, we conclude that the coverage rates of the confidence intervals are all close to its nominal truth 0.95 under different settings. One possible reason for this phenomenon is that the proposed estimator is model free, so the proposed model is more robust and can be used under complex sampling designs. 

\begin{table}[!ht]
	\footnotesize
	\centering% This table is typed based the previous result.
	\caption{Coverage rate of a 95\% confidence interval by the proposed method based on 1,000 Monte Carlo simulations under different setups.  ``Homo'' and ``Hete'' stands for the homogeneous and heterogeneous variance for the error term, respectively. ``$\mathcal{S}_1$ Des'' and ``$\mathcal{S}_2$ Des'' show the sampling design used to generate the internal sample $\mathcal{S}_1$  and the external sample $\mathcal{S}_2$. ``SRS'' and ``Poi'' stands for SRS and Poisson sampling, respectively. ``Independent'' and ``Dependent'' correspond to the cases when the auxiliary variables are independent and dependent, respectively.  }
	\begin{tabular}{cccccccccc}
		\hline
		&\multirow{2}{*}{$\mathcal{S}_1$ Des}&\multirow{2}{*}{$\mathcal{S}_2$ Des}&\multicolumn{3}{c}{Independent}&&\multicolumn{3}{c}{Dependent}\\
		\cline{4-6}\cline{8-10}
		&&&$\beta_0$&$\beta_1$&$\beta_2$&&$\beta_0$&$\beta_1$&$\beta_2$\\
		\hline
		\multirow{5}{*}{Homo}&\multirow{2}{*}{SRS}&SRS &0.948&0.952&0.939&&0.945&0.948&0.934\\
		&    &Poi &0.945&0.951&0.938&&0.946&0.946&0.934\\\vspace{-1em}\\
		&\multirow{2}{*}{Poi}&SRS &0.957&0.966&0.949&&0.935&0.943&0.940\\
		&    &Poi &0.962&0.964&0.951&&0.936&0.943&0.938\\\vspace{-1em}\\\\
		\multirow{5}{*}{Hete}&\multirow{2}{*}{SRS}&SRS &0.944&0.942&0.933&&0.933&0.925&0.935\\
		&    &Poi &0.949&0.942&0.935&&0.935&0.934&0.931\\\vspace{-1em}\\
		&\multirow{2}{*}{Poi}&SRS &0.959&0.955&0.935&&0.948&0.950&0.941\\
		&    &Poi &0.961&0.956&0.944&&0.952&0.949&0.946\\
		\hline                          
	\end{tabular}
	
	\label{tab: SRS CI}
\end{table}

An additional simulation with a logistic regression setup is relegated to  \ref{ss: logistic regression} of the Supplementary Material, and similar conclusions can be reached.

\section{Application Study}
\label{sec: application}
\subsection{Data Description and Problem Formulation}

As an application example, we apply the proposed method to analyze a subset of the data from the Korea National Health and Nutrition Examination Survey (KNHANES). The annual survey includes approximately 5{,}000 individuals each year and collects information regarding health-related behaviors by interviews, basic health conditions by physical and blood tests, and dietary intake by nutrition survey. The sampling design of KNHANES is a stratified sampling using age, sex, and region as stratification variables. The final sampling weights are computed via nonresponse adjustment and post-stratification, then provided to data users with survey variables. 

To improve the efficiency of data analysis with KNHANES of size $n_1=4{,}929$, we used an external public database provided by the National Health Insurance Sharing Service (NHISS) in Korea. The big data provided by NHISS contain about $n_2=1{,}000{,}000$  individuals with health-related information, some of whose variables are a subset of variables in KNHANES. %\textcolor{red}{(Please provide a table describing the available items for two data sources.) }

These data structures, with the small $n_1$, the large $n_2$, and the big data having a subset of variables in the internal sample, are suited well to the setting we addressed in Section 2. 
However, there is another complication in applying the proposed method to the real application. In the NHISS data, its selection probabilities are unknown, so the design consistent estimator $\hat{\bmalpha}_2$ in \eqref{gls} is unavailable. Section \ref{sec:prop_weigting} addresses this issue by using a propensity weighting approach and Section \ref{sec:application_result} presents the analysis result of the application study. 

\subsection{Propensity Weighing for External Data with Unknown Selection Probability}
\label{sec:prop_weigting}

We now consider an extension of the proposed method to the case where the external sample $\mathcal{S}_2$ is a big data with unknown selection probabilities. In this case, the working model for $E(Y_i \mid \bmx_{i1}) = m( \bmalpha^\top \bmx_{i1})$ may not hold for the sample $S_2$.  Nonetheless, we may still solve 
\begin{eqnarray} 
	\sum_{i \in \mathcal{S}_2} \{ y_i - m(\bmalpha^\top \bmx_{i1} ) \} \bmx_{i1} = \bmzero
	\label{eq: s1}
\end{eqnarray} 
to obtain $\hat{\alpha}_0$ and $\hat{\alpha}_1$. If the sampling mechanism for $\mathcal{S}_2$ is ignorable or non-informative, then the solution of \eqref{eq: s1} is unbiased; otherwise, the resulting estimator is biased. 

To remove the selection biases in the big data estimate, 
\cite{kimwang2020} suggested using  propensity score weights in \eqref{eq: s1} to obtain an unbiased estimator of $\bmalpha$.  
To construct the propensity score  weights, we employ a nonignorable nonresponse model, 
%  \begin{equation} 
$P( \delta_i = 1 \mid \bmx_{i1}, y_i) = \pi( \bmx_{i1}, y_i; \bmphi)$, 
%  \label{smooth}\notag
%  \end{equation} 
%  
where $\delta_i = 1$ if $i \in S_2$ and zero otherwise. 
Note that we can express 
%  \begin{equation}
%  \frac{1}{\pi (\bmx_{i1}, y_i) } = 1 + \frac{N_0}{N_1} r(\bmx_{i1}, y_i) 
%  \end{equation}
$\pi (\bmx_{i1}, y_i)^{-1} = 1 + (N_0/N_1) r(\bmx_{i1}, y_i)$
where 
$r(\bmx_{i1}, y_i) = f(\bmx_{i1}, y_i | \delta_i = 0 ) / f (\bmx_{i1}, y_i | \delta_i = 1)$
is the density ratio function with $N_1= \sum_{i=1}^N \delta_i $ and $N_0 = N - N_1$.  
Using the motivation of \cite{wang2021}, we may assume a log-linear density ratio model,  
%  \begin{equation} 
$\log \{ r(x_{i1}, y_i; \bmphi) \} = \phi_0 + \phi_1 x_{i1} + \phi_2 y_i$. 
%   \label{model1}\notag
%   \end{equation} 
The maximum entropy estimator of $\bmphi$ is obtained by solving 
%  $$ \frac{1}{N_1}  \sum_{i=1}^N  \delta_i  \exp (\phi_0 +  \phi_1 x_{i1} + \phi_2 y_i ) (1, x_{i1}, y_i)   = \left(1, \hat{\bar{x}}_1, \hat{\bar{y}} \right) ,
%  $$
$(1/N_1)  \sum_{i=1}^N  \delta_i  \exp (\phi_0 +  \phi_1 x_{i1} + \phi_2 y_i ) (1, x_{i1}, y_i) = \left(1, \hat{\bar{x}}_1, \hat{\bar{y}} \right)$
where  
$(\hat{\bar{x}}_1, \hat{\bar{y}}) = 
(1/\hat{N}_0) \left\{ \sum_{i \in \mathcal{S}_1} d_i  (x_{i1}, y_i)- \sum_{i=1}^N  \delta_i (x_{i1}, y_i) 
\right\}
$
and $\hat{N}_0 = \sum_{i \in S_1} d_i - N_1$ where $\mathcal{S}_1$ is the internal sample. 
Once $\hat{\bmphi}$ is obtained, we can construct $\hat{\pi}(x_{i1}, y_i)$ and solve 
\begin{eqnarray} 
	\sum_{i \in \mathcal{S}_2} \frac{1}{ \hat{\pi} (x_{i1}, y_i)} \{ y_i - m(\alpha_0+ \alpha_1 x_{i1} ) \} (1, x_{i1} ) = (0, 0) 
	\label{s2}
\end{eqnarray} 
to obtain $\hat{\bmalpha}_2 = (\hat{\alpha}_0, \hat{\alpha}_1)$.

In addition, we can use the internal sample $\mathcal{S}_1$ to fit the same working model to obtain $\hat{\bmalpha}_1$. After that, we obtain 
%  \begin{equation} 
%  \hat{\bmalpha}^* = \frac{ \bmV^{-1}  \hat{\bmalpha} + \bmV_2^{-1} \hat{\bmalpha}_2 }{ \bmV^{-1} + \bmV_2^{-1}} 
% \label{gls1} 
%  \end{equation} 
$\hat{\bmalpha}^*$ using \eqref{gls} 
and apply the proposed calibration weighting method to combine information from the big data. In practice, $\bmV_2$ in \eqref{gls} is difficult to compute, but it is negligibly small if the sample size for $S_2$ is huge. In this case, we may simply use $\hat{\bmalpha}^*=\hat{\bmalpha}_2$ in  the calibration problem.

\subsection{Application Study Results: Korea National Health and Nutrition Examination Survey}
\label{sec:application_result}

In this application study, we use $n_1=4{,}929$ records of KNHANES data that have no missing values in four variables: Total cholesterol, Hemoglobin, Triglyceride, and high-density lipoprotein (HDL) cholesterol. For demonstration purpose, we assume that an analyst is interested in conducing the following linear regression analysis,
$$ E( \text{Total Cholesterol}_i | \bmx_i ) = \beta_0 + \beta_1 \text{Hemoglobin}_i + \beta_2 \text{Triglyceride}_i + \beta_3 \text{HDL}_i \BB \text{for } i \in \mathcal{S}_1;$$
check Section~\ref{ss: valid linearity} of the Supplementary Material for  details about the linearity assumption.
In our data, the biggest absolute value of the pairwise correlation among covariates is -0.40 observed between Triglyceride and HDL cholesterol, which is similar to a scenario in Section \ref{sec:simulation_study} where the covariates were highly correlated. 
% Correlation among continuous variables 
% {\small
% \begin{verbatim}
%         HGB    TG TCHOL   HDL
% HGB    1.00  0.27  0.12 -0.19
% TG     0.27  1.00  0.28 -0.40
% TCHOL  0.12  0.28  1.00  0.18
% HDL   -0.19 -0.40  0.18  1.00
% \end{verbatim}
% }
The big external data consist of $n_2 = 1,000,000$ records of NHISS data with fully observed items in Total cholesterol, Hemoglobin, and Triglyceride. The assumed working reduced model  is 
$$ E( \text{Total Cholesterol}_i | \bmx_{i1} ) = \alpha_0 + \alpha_1 \text{Hemoglobin}_i + \alpha_2 \text{Triglyceride}_i \BB \text{for } i \in \mathcal{S}_1 \cup \mathcal{S}_2. $$

In this application study, we implement our proposed methods with the external sample where $\hat{\bmalpha}_2$ is used instead of $\bmalpha^*$ that is unavailable as we do not have information regarding the entire population. With the external sample whose selection probabilities are unknown, we prepare two versions of proposed methods: (i) considering $\mathcal{S}_2$ as SRS, i.e., without propensity weighting, and (ii) with the propensity weighting adjustment introduced in Section \ref{sec:prop_weigting}. 
For the propensity weighting, we fit the log-linear density ratio model to the external data, $\log \{ r(\bmx_{i1}, y_i; \bmphi) \} = \phi_0 + \phi_1 \text{Hemoglobin}_i + \phi_2 \text{Triglyceride}_i + \phi_3 \text{Total Cholesterol}_i$, calculate $\hat{\pi}(\bmx_{i1}, y_i)$ given $\hat{\bmphi}$, then solve \eqref{s2} to obtain $\hat{\bmalpha}_2$. The above logistic regression model is commonly assumed in the literure; see \citet{elliott2017inference}, \citet{chen2020doubly}, \citet{wang2021} and the references within for details. {Since the CML estimator fails to incorporate the design features, it is not considered in the application section.}
The performances of proposed methods are compared with the reference method that uses the internal sample $\mathcal{S}_1$ only to get weighted least square estimates considering the sampling weights. 

\begin{figure}
	\begin{center}
		\includegraphics[width=1.0\textwidth]{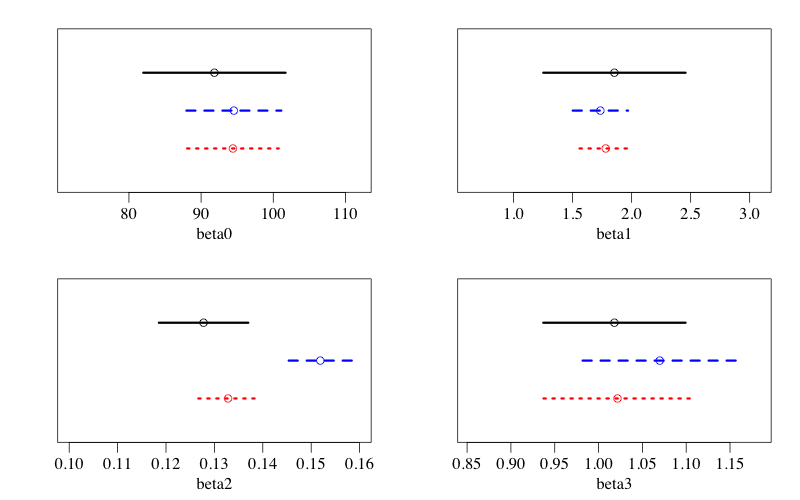}
	\end{center}
	\caption{Comparison of the regression analysis for $E( \text{Total Cholesterol}_i | \bmx_i ) = \beta_0 + \beta_1 \text{Hemoglobin}_i + \beta_2 \text{Triglyceride}_i + \beta_3 \text{HDL}_i$ using the internal data from Korea National Health and Nutrition Examination Survey supported by the big external data from the National Health Insurance Sharing Service database. For each panel, circles are point estimates and lines are their 95\% confidence intervals for using the internal sample $\mathcal{S}_1$ only with the weighted least square (top solid line), the proposed method without adjustment (middle dashed line), and the proposed method with propensity score weighting adjustment (bottom dotted line).  \label{fig:app_result}}
\end{figure}

Figure \ref{fig:app_result} shows the point estimates and the 95\% confidence intervals of $\hat{\bmbeta}=(\hat{\beta}_0,\hat{\beta}_1,\hat{\beta}_2,\hat{\beta}_3)$ for each method. The proposed methods show smaller variances for $\hat{\beta}_0$, $\hat{\beta}_1$ and $\hat{\beta}_2$ than using the internal sample only. This result coincides with our findings in the simulation studies of the previous section. For $\beta_2$, the estimator of the proposed method without propensity weighting shows a systematic difference from the other two estimators. When the propensity weighting adjustment is coupled with the proposed method, its confidence interval of $\beta_2$ is contained by that of using the internal sample only. This result implies that the systematic bias due to the disregard of the sampling probabilities is addressed by the propensity weighting adjustment. No efficiency gain in estimating $\beta_3$ was expected as the external data contain information of $x_{i1}$ (Hemoglobin) and $x_{i2}$ (Triglyceride), not $x_{i3}$ (HDL). 

% {\bf Results: Points estimates (and SE)}

% \bigskip 
% \begin{tabular}{lcccc}
% \hline 
% &   & hemoglobin & triglyceride & HDL \\
% & $\beta_0$ & $\beta_1$ & $\beta_2$ & $\beta_3$ \\ 
% \hline 
% Using Internal Survey only - Weighted & 91.8 & 1.85 & 0.13 & 1.02 \\
% & (5.23) & (0.32) & (0.007) & (0.043) \\
% Proposed WO propensity adj & 94.5 & 1.74 & 0.15 & 1.07 \\
% & (3.35) & (0.12) & (0.003) & (0.045) \\
% Proposed with propensity adj & 94.4 & 1.78 & 0.13 & 1.02 \\
% & (3.24) & (0.11) & (0.003) & (0.043) \\
% \hline 
% \end{tabular}

% \bigskip 
% Variance Estimation for the WLS (using package `survey')

% \noindent {\bf Comparison of $\hat{\alpha}$'s from $S_A$ (the survey data) and $S_2$ (the big external data)}

% \bigskip 
% \begin{tabular}{lccc}
% \hline 
% &   & hemoglobin & triglyceride \\
% & $\alpha_0$ & $\alpha_1$ & $\alpha_2$ \\ 
% \hline 
% Using internal survey - OLS & 159.9 & 1.14 & 0.086 \\
% & (4.39) & (0.32) & (0.005) \\
% Using internal survey - Weighted & 159.7 & 1.15 & 0.085 \\
% & (4.56) & (0.33) & (0.005) \\
% Using External data & 166.8 & 1.01 & 0.104 \\
% & (0.31) & (0.02) & ($<$0.001) \\
% \hline 
% \end{tabular}

% \noindent Check that $\hat{\alpha}$'s from the survey data and the big data are different
% $\Rightarrow$ If they are similar, the sampling design of the big data is not informative. $\Rightarrow$ Then, we do not need the propensity score adjustment proposed in this document. 

%\newpage
\section{Conclusion}
\label{sec:conc}

Incorporating external data sources into the regression analysis of the internal sample is an important practical problem. We have addressed this problem using a novel application of the information projection \citep{csiszar2004} and 
the model calibration weighting  \citep{wu2001model}. The proposed method is directly applicable to survey sampling and can be easily extended to multiple data integration. The proposed method is easy to implement and does not require direct access to external data. As long as the estimated regression coefficients and their  standard errors for the working reduced model are available, we can incorporate the extra information into our analysis. 

There are several possible directions on future research extensions. First, a Bayesian approach can be developed under the same setup. One may use the Bayesian empirical likelihood method of \citet*{zhao2020} in this setup.   The proposed method can potentially be used to combine the randomized clinical trial data with  big real-world data \citep{yang2020b}; such extensions will be presented elsewhere.  
It will be also interesting to connect the proposed approach to two-phase (double) sampling design whose efficient design and estimation has been recently studied actively \citep*{rivera2019analysis,rivera2020augmented,wang2020novel}. The data structure of the two-phase sampling with the large-$n$, small-$p$ first stage sample and the small-$n$, large-$p$ second stage sample is well suited to the set-up assumed by the suggested model calibration approach. 

\section*{Acknowledgements}
We appreciate the constructive comments of the reviewers and the AE. 
The research of Z. Wang was partially supported by the National Natural Science Foundation of China grants (Award no: 11901487, 72033002) and the Fundamental Scientific Center of National Natural Science Foundation of China grant (Award no: 71988101). The research of J.K. Kim was partially supported by the National Science Foundation grant (Award no: CSSI-1931380), a Cooperative Agreement between the
US Department of Agriculture Natural Resources Conservation Service and Iowa State University,  and the Iowa Agriculture and Home Economics Experiment Station,
Ames, Iowa.

\appendix
	\setcounter{theorem}{0}
	\setcounter{lemma}{0}
\setcounter{equation}{0}	
\setcounter{table}{0}	
\setcounter{figure}{0}	
\renewcommand{\theequation}{A.\arabic{equation}}
	\renewcommand{\thelemma}{A\arabic{lemma}}
	\renewcommand{\thetheorem}{A\arabic{theorem}}
	\renewcommand{\thetable}{A\arabic{table}}
\renewcommand{\thefigure}{A\arabic{figure}}
\section*{Appendix}
	\section{Proof of Theorem~\ref{theo: CLT}}\label{proof: CLT}
		\begin{lemma}\label{theo: lambda cons}
	Suppose that conditions C\ref{cond: max U1i}\removed{, C\ref{cond: cond2 hat N}}, C\ref{cond: variance conv} and C\ref{cond: CLT} hold. Then, $\lVert\hat{\bmlambda}\rVert = O_p(n^{-1/2})$.
    \end{lemma}
	\begin{proof}[Proof of Lemma~\ref{theo: lambda cons}]
		Denote $\hat{\blambda} = \rho\btheta$, where $\rho = \lVert\hat{\blambda}\rVert$ and $\btheta = \rho^{-1}\hat{\blambda}$ is a vector of unit length. Then, we have 
		\begin{eqnarray}
			\bmzero &=&\left\lvert \sum_{i\in \mathcal{S}_1}\frac{\tilde{d}_i}{1-\hat{\blambda}^{\T}\bmU_2(\balpha_N^*;\bmx_{1i},y_i) }\bmU_2(\balpha_N^*;\bmx_{1i},y_i) \right\rvert\notag \\ 
			&=& \left\lvert \btheta^{\T}\sum_{i\in \mathcal{S}_1}\frac{\tilde{d}_i}{1-\rho\btheta^{\T}\bmU_2(\balpha_N^*;\bmx_{1i},y_i) }\bmU_2(\balpha_N^*;\bmx_{1i},y_i) \right\rvert\notag\\
			&=&\left\lvert \sum_{i\in \mathcal{S}_1}\tilde{d}_i\btheta^{\T}\bmU_2(\balpha_N^*;\bmx_{1i},y_i)  +\rho\sum_{i\in \mathcal{S}_1}\frac{\tilde{d}_i\btheta^{\T}\bmU_2(\balpha_N^*;\bmx_{1i},y_i) \{\bmU_2(\balpha_N^*;\bmx_{1i},y_i)\} ^{\T}\btheta}{1-\rho\btheta^{\T}\bmU_2(\balpha_N^*;\bmx_{1i},y_i) }\right\rvert\notag \\ 
			&\geq& \frac{\rho}{1+\rho Z_S}\left\lvert\sum_{i\in \mathcal{S}_1}\tilde{d_i}\btheta^{\T}\bmU_2(\balpha_N^*;\bmx_{1i},y_i) \{\bmU_2(\balpha_N^*;\bmx_{1i},y_i)\} ^{\T}\btheta \right\rvert - \left\lvert\sum_{i\in \mathcal{S}_1}\tilde{d}_i\btheta^{\T}\bmU_2(\balpha_N^*;\bmx_{1i},y_i) \right\rvert,\notag \\ 
			\label{eq: theorem 1 1}
		\end{eqnarray}
		where  the first equality holds since $g(\hat{\beeta}) = \bmzero$, the last inequality holds by the triangular inequality. 
		
		By C\ref{cond: variance conv} and the  Rayleigh-Ritz Theorem \citep[Section~4.2]{horn2017}, there exists a constant $\sigma_0>0$ such that 
		\begin{equation}
			\sum_{i\in \mathcal{S}_1}\tilde{d}_i\btheta^{\T}\bmU_2(\balpha_N^*;\bmx_{1i},y_i)\{\bmU_2(\balpha_N^*;\bmx_{1i},y_i)\}^{\T}\btheta >\sigma_0+o_p(1).
		\end{equation}
		By C\ref{cond: CLT} and the Slutsky's theorem, we have 
		\begin{eqnarray}
			&\sum_{i\in \mathcal{S}_1}\tilde{d}_i\btheta^{\T}\bmU_2(\balpha_N^*;\bmx_{1i},y_i) = O_p(n^{-1/2}).\label{eq: theorem 1 2}
		\end{eqnarray}
		Thus, by C\ref{cond: max U1i} and (\ref{eq: theorem 1 1})--(\ref{eq: theorem 1 2}), we have proved Lemma~\ref{theo: lambda cons}.
	\end{proof}
% 	\section{Proof of Lemma~\ref{theo: beta cons}}\label{proof: beta constant}   
	\begin{lemma}\label{theo: beta cons}
	Suppose that conditions C\ref{cond: max U1i}, \removed{C\ref{cond: cond2 hat N}}C\ref{cond: beta region}--C\ref{cond: variance conv} and C\ref{cond: CLT}  hold. Then, $\hat{\bmbeta}-\bmbeta_{0N}=o_p(1)$.
    \end{lemma}
	\begin{proof}[Proof of Lemma~\ref{theo: beta cons}]
		By Lemma~~\ref{theo: lambda cons} and C\ref{cond: max U1i}, we conclude that 
		\begin{equation}
			\max_{i\in \mathcal{S}_1}\lvert \hat{\blambda}^{\T}\bmU_2(\balpha_N^*;\bmx_{1i},y_i)\rvert\leq \max_{i\in \mathcal{S}_1}\lVert \hat{\blambda}\rVert \lVert \bmU_2(\balpha_N^*;\bmx_{1i},y_i)\rVert =\lVert \hat{\blambda}\rVert \max_{i\in \mathcal{S}_1}\lVert \bmU_2(\balpha_N^*;\bmx_{1i},y_i)\rVert =  o_p(1).\label{eq: theorem 2 max lambdau}
		\end{equation}
		
		First, we show that 
		\begin{equation}
			\sum_{i\in \mathcal{S}_1}\frac{\tilde{d}_i}{1-\hat{\blambda}^{\T}\bmU_2(\balpha_N^*;\bmx_{1i},y_i)}\bmU_1(\bmbeta;\bmx_i,y_i)  - \sum_{i\in \mathcal{S}_1}\tilde{d}_i\bmU_1(\bmbeta;\bmx_i,y_i)\to \bmzero \label{eq: theorem2 1}
		\end{equation}
		in probability uniformly for  $\bbeta\in \Omega$.
		By (\ref{eq: theorem 2 max lambdau}), we have 
		\begin{eqnarray}
			&&\left\lVert\sum_{i\in \mathcal{S}_1}\tilde{d}_i\left\{\frac{1}{1-\hat{\blambda}^{\T}\bmU_2(\balpha_N^*;\bmx_{1i},y_i)}-1\right\}\bmU_1(\bmbeta;\bmx_i,y_i) \right\rVert \notag \\ 
			&=& \left\lVert\sum_{i\in \mathcal{S}_1}\tilde{d}_i\{ \hat{\blambda}^{\T}\bmU_2(\balpha_N^*;\bmx_{1i},y_i)+o_p(\hat{\blambda}^{\T}\bmU_2(\balpha_N^*;\bmx_{1i},y_i))\}\bmU_1(\bmbeta;\bmx_i,y_i) \right\rVert\notag \\
			&\leq&(1+o_p(1))\max_{i\in \mathcal{S}_1}\lvert \hat{\blambda}^{\T}\bmU_2(\balpha_N^*;\bmx_{1i},y_i)\rvert\left\lVert\sum_{i\in \mathcal{S}_1}\tilde{d}_i\bmU_1(\bmbeta;\bmx_i,y_i) \right\rVert.\label{eq: difference abs}
		\end{eqnarray}
		By C\ref{cond: beta region}--C\ref{cond: uniform abs}, there exists a constant $C_{u1}>0$ such that
		$\sup_{\boldsymbol{\beta}\in\Omega}\lVert\bmU_{0}(\bbeta)\rVert<C_{u1}$. Since $\sum_{i\in \mathcal{S}_1}\tilde{d}_i \bmU_1(\bmbeta;\bmx_i,y_i)$ converge uniformly to $\bmU_{0}(\bbeta)$ in probability, we conclude that 
		\begin{equation}
			\left\lVert\sum_{i\in \mathcal{S}_1}\tilde{d}_i\bmU_1(\bmbeta;\bmx_i,y_i) \right\rVert<C_{u1}+o_p(1)\label{eq: theorem 2 second part}
		\end{equation}
		uniformly over $\Omega$. 
		By (\ref{eq: theorem 2 max lambdau}) and (\ref{eq: difference abs})--(\ref{eq: theorem 2 second part}), we have validated (\ref{eq: theorem2 1}). 
		
		By C\ref{cond: unique beta} and (\ref{eq: theorem2 1}), we conclude that $\sum_{i\in \mathcal{S}_1}\tilde{d}_i\{1-\hat{\blambda}^{\T}\bmU_2(\balpha_N^*;\bmx_{1i},y_i)\}^{-1}\bmU_1(\bmbeta;\bmx_i,y_i) $ converges uniformly to $U_0(\bbeta)$ in probability. Denote $Q_0(\bbeta)=-U_0(\bbeta)^2$ and $Q_{s}(\bbeta) = -[\sum_{i\in \mathcal{S}_1}\tilde{d}_i\{1-\hat{\blambda}^{\T}\bmU_2(\balpha_N^*;\bmx_{1i},y_i)\}^{-1}\bmU_1(\bmbeta;\bmx_i,y_i) ]^2$. Then, $\bbeta_p$ uniquely maximizes $Q_0(\bbeta)$ by (C\ref{cond: unique beta}), and $\hat{\bbeta}$ maximizes $Q_{s}(\bbeta)$. In addition,  $Q_{s}(\bbeta)$ converge uniformly to $Q_0(\bbeta)$ in probability over the compact set $\Omega$. Thus, by C\ref{cond: beta region} and Theorem~2.1 of \citet[Chapter~36]{engle1994handbook}, we have finished the proof for Lemma~\ref{theo: beta cons}.        
	\end{proof}
	\begin{proof}[Proof of Theorem~\ref{theo: CLT}] By Lemmas~\ref{theo: lambda cons}--\ref{theo: beta cons}, we have shown that 
		\begin{equation}
			\hat{\beeta} = \beeta_0 + o_p(1),\label{eq: theorem 3 1}
		\end{equation}
		where $\hat{\beeta}^{\T} = (\hat{\bbeta}^{\T},\hat{\blambda}^{\T})$, $\beeta_0^{\T} = ( \bbeta_{0N}^{\T},\bzero^{\T})$, and $\bzero$ is a vector of zero with the same length of $\hat{\blambda}$.
		
		By (\ref{eq: theorem 3 1}) and the Taylor expansion, we have 
		\begin{eqnarray}
			\bmzero &=& g(\hat{\beeta}) = g(\beeta_0) + \frac{\partial g}{\partial \beeta^{\T}}(\beeta_0)(\hat{\beeta} - \beeta_0) + o_p(\lVert \hat{\beeta} - \beeta_0\rVert)\notag \\
			&=& \begin{pmatrix}
				\displaystyle\sum_{i\in \mathcal{S}_1}\tilde{d}_i \bmU_1(\bmbeta_{0N};\bmx_i,y_i)\\\displaystyle\sum_{i\in \mathcal{S}_1}\tilde{d}_i \bmU_2(\balpha_N^*;\bmx_{1i},y_i)
				
			\end{pmatrix}  \notag \\ 
			&&+ 
			\begin{pmatrix}
				\displaystyle\sum_{i\in \mathcal{S}_1}\tilde{d}_i\frac{\partial \bmU_1(\bmbeta_{0N};\bmx_i,y_i)}{\partial\bbeta^{\T}} &\displaystyle\sum_{i\in \mathcal{S}_1}\tilde{d}_i\bmU_1(\bmbeta_{0N};\bmx_i,y_i)\{\bmU_2(\balpha_N^*;\bmx_{1i},y_i)\}^{\T}\\\bmzero&\displaystyle\sum_{i\in \mathcal{S}_1}\tilde{d}_i\{\bmU_2(\balpha_N^*;\bmx_{1i},y_i)\}^{\otimes2}
				
			\end{pmatrix}(\hat{\beeta} - \beeta_0)\notag \\ 
			&&+o_p(\lVert \hat{\beeta} - \beeta_0\rVert).\notag \\\label{eq: theorem 3 CLT 1} 
		\end{eqnarray}
		
		By \removed{(C\ref{cond: cond2 hat N}) and} (C\ref{cond: CLT}), we have 
		\begin{equation}
			n^{1/2}\begin{pmatrix}
				\displaystyle\sum_{i\in \mathcal{S}_1}\tilde{d}_i \bmU_1(\bmbeta_{0N};\bmx_i,y_i)\\
				\displaystyle\sum_{i\in \mathcal{S}_1}\tilde{d}_i \bmU_2(\balpha_N^*;\bmx_{1i},y_i)
				
			\end{pmatrix} \to \mathcal{N}(\bmzero,\bmSigma_u)
		\end{equation}
		in distribution. By (C\ref{cond: variance conv})--(C\ref{cond: I 22}), we conclude that
		\begin{equation}
			\begin{pmatrix}
				\displaystyle\sum_{i\in \mathcal{S}_1}\tilde{d}_i\frac{\partial \bmU_1(\bmbeta_{0N};\bmx_i,y_i)}{\partial\bbeta^{\T}}& \displaystyle\sum_{i\in \mathcal{S}_1}\tilde{d}_i\bmU_1(\bmbeta_{0N};\bmx_i,y_i)\{\bmU_2(\balpha_N^*;\bmx_{1i},y_i)\}^{\T}\\
				\bmzero& \displaystyle\sum_{i\in \mathcal{S}_1}\tilde{d}_i\{\bmU_2(\balpha_N^*;\bmx_{1i},y_i)\}^{\otimes2}
			\end{pmatrix}\to\bmcalI\label{eq: theorem 3 CLT 3}
		\end{equation}
		in probability, where 
		$$
		\bmcalI = \begin{pmatrix}
			\bmcalI_{11}&\bmcalI_{12}\\
			\bmzero&\bmcalI_{22}
		\end{pmatrix}.
		$$
		By (\ref{eq: theorem 3 CLT 1})--(\ref{eq: theorem 3 CLT 3}), we conclude that 
		\begin{equation}
			n^{1/2}(\hat{\beeta} - \beeta_0)\to \mathcal{N}(\bmzero,\bmSigma_\eta)
		\end{equation}
		in distribution, 
		where $
		\bmSigma_\eta = \bmcalI^{-1}\bmSigma_u(\bmcalI^{-1})^{\T}.
		$
	\end{proof}
	
		\section{Proof of Corollary~\ref{cor: optimal var}}\label{sec app proof cor opti}
	Since $\bmU_1(\bmbeta;\bmx,y)$ is given, it is enough to consider 
	$$
	\bmSigma_{11} -  \bmcalI_{12}\bmcalI_{22}^{-1} \bmSigma_{21} - \bmSigma_{12}\bmcalI_{22}^{-1} \bmcalI_{12}^{\T}
	+\bmcalI_{12}\bmcalI_{22}^{-1} \bmSigma_{22}\bmcalI_{22}^{-1} \bmcalI_{12}^{\T}, 
	$$
	the asymptotic variance of $\tilde{\bmU}_1(\bmbeta_{0N}) - \bmcalI_{12}\bmcalI_{22}^{-1}\tilde{\bmU}_2(\bmalpha_N^*)$, where $\tilde{\bmU}_1(\bmbeta_{0N}) = n^{1/2}\sum_{i\in\mathcal{S}_1}\tilde{d}_i\bmU_1(\bmbeta_{0N};\bmx_i,y_i)$ and $\tilde{\bmU}_2(\bmalpha_N^*) = n^{1/2}\sum_{i\in\mathcal{S}_1}\tilde{d}_i\bmU_2(\bmalpha_N^*;\bmx_{i1},y_i)$.
	
	Consider 
	\begin{eqnarray}
		&&\var\{\tilde{\bmU}_1(\bmbeta_{0N}) - \bmcalI_{12}\bmcalI_{22}^{-1}\tilde{\bmU}_2(\bmalpha_N^*)\} \notag \\ 
		&=&E[\var\{\tilde{\bmU}_1(\bmbeta_{0N}) - \bmcalI_{12}\bmcalI_{22}^{-1}\tilde{\bmU}_2(\bmalpha_N^*)\mid \mathcal{A}_N\}] + \var[E\{\tilde{\bmU}_1(\bmbeta_{0N}) - \bmcalI_{12}\bmcalI_{22}^{-1}\tilde{\bmU}_2(\bmalpha_N^*)\mid \mathcal{A}_N\}]\notag \\
		&\succcurlyeq&E[\var\{\tilde{\bmU}_1(\bmbeta_{0N})\}],\notag
	\end{eqnarray}
	where $\mathcal{A}_N = \{(\bmx_{i1},y):i\in\mathcal{S}_1\}$, $A\succcurlyeq B$ is equivalent to that $A-B$ is non-negatively definitive for two matrices $A$ and $B$ with the same dimension, the last inequality holds since $\tilde{\bmU}_2(\bmalpha_N^*)$ is non-stochastic conditional on $\mathcal{A}_N$. Thus, $\var\{\tilde{\bmU}_1(\bmbeta_{0N}) - \bmcalI_{12}\bmcalI_{22}^{-1}\tilde{\bmU}_2(\bmalpha_N^*)\}$ achieves minimum if $\var[E\{\tilde{\bmU}_1(\bmbeta_{0N}) - \bmcalI_{12}\bmcalI_{22}^{-1}\tilde{\bmU}_2(\bmalpha_N^*)\mid \mathcal{A}_N\}]=0$, which is induced by the condition $\bmcalI_{12}\bmcalI_{22}^{-1}\bmU_2(\bmalpha_N^*;\bmx_{1},y) = E\{\bmU_1(\bmbeta_{0N};\bmx,y)\mid \bmx_{1},y\}$.

	\section{Proof of Theorem~\ref{theo: CLT hat alpha *}}\label{SS: PROOF T4}
	\setcounter{lemma}{2}
	Before proving  Theorem~\ref{theo: CLT hat alpha *}, we need the following result.
	\begin{lemma}\label{lemma: order of alpha*}
		Suppose that conditions C\ref{cond: max U1i}, C\ref{cond: CLT}--C\ref{cond: consistent alphas} hold. Then, we have
		$$
		\hat{\balpha}^* - \balpha^*_{N} = O_p(n^{-1/2}),
		$$
	\end{lemma}
	\begin{proof}[Proof of Lemma~\ref{lemma: order of alpha*}]
		Since $\hat{\balpha}_{2}$ is obtained by an independent external survey, we conclude that the variance of $\hat{\balpha}^*$ can be estimated by $(\bmV_1^{-1} + \bmV_{2}^{-1})^{-1}$. Thus, the order of the variance of $\hat{\balpha}^*$ is determined by the less efficient estimator between $\hat{\balpha}_1$ and $\hat{\balpha}_{2}$. If we showed
		\begin{equation}
			\hat{\balpha}_1 - \balpha^*_{N} = O_p(n^{-1/2}), \label{eq: order of alpha1}
		\end{equation}
		we could have $\bmV_1 = O_p(n^{-1})$ by (C\ref{cond: var consist alphas}). Since $ \hat{\balpha}^*$ is at least as efficient as $\hat{\balpha}_1$,  we have completed the proof of Lemma~\ref{lemma: order of alpha*}. 
		
		Thus, it remains to show (\ref{eq: order of alpha1}). 
		By C\ref{cond: I alpha} and C\ref{cond: consistent alphas}, we have 
		\begin{eqnarray}
			\bmzero &=& \sum_{i\in \mathcal{S}_1}\tilde{d}_i\bmU_2(\hat{\balpha}_1;\bmx_{1i},y_i)\notag \\ 
			&=& \sum_{i\in \mathcal{S}_1}\tilde{d}_i\bmU_2(\balpha_N^*;\bmx_{1i},y_i) + \left\{\frac{\partial}{\partial\balpha^{\T}}\sum_{i\in \mathcal{S}_1}\tilde{d}_{i}\bmU_2(\tilde{\bmalpha};\bmx_{1i},y_i)\right\}(\hat{\bmalpha}_1 - \balpha^*_{N}),\notag \\ 
			\label{eq: U1 alpha 1 =0}
		\end{eqnarray}
		where $\tilde{\balpha}$ lies on the segment joining $\hat{\bmalpha}_1$ and  $\balpha^*_{N}$. 
		By C\ref{cond: CLT}--C\ref{cond: consistent alphas} and \eqref{eq: U1 alpha 1 =0}, we conclude that
		\begin{eqnarray}
			\hat{\balpha} - \balpha^*_{N} = -\bmcalI_{0}^{-1}\sum_{i\in \mathcal{S}_1}\tilde{d}_i\bmU_2(\balpha_N^*;\bmx_{1i},y_i)+o_p(n^{-1/2}).\label{eq: alpha1 - alpha*}
		\end{eqnarray}
		Thus, by C\ref{cond: CLT} and (\ref{eq: alpha1 - alpha*}), we have shown (\ref{eq: order of alpha1}).
	\end{proof}
	
	\begin{proof}[Proof of Theorem~\ref{theo: CLT hat alpha *}]
		Consider
		\begin{eqnarray}
			&&\sum_{i\in \mathcal{S}_1}\tilde{d}_{i}\bmU_2(\hat{\balpha}^*;\bmx_{1i},y_i)\notag \\  
			&=&  \sum_{i\in \mathcal{S}_1}\tilde{d}_{i}\bmU_2(\balpha_N^*;\bmx_{1i},y_i) + \left\{\frac{\partial}{\partial\balpha^{\T}}\sum_{i\in \mathcal{S}_1}\tilde{d}_{i}\bmU_2(\tilde{\balpha};x_{1i},y_i)\right\}(\hat{\balpha}^* - \balpha_N^*)\notag \\
			&=&  \sum_{i\in \mathcal{S}_1}\tilde{d}_{i}\bmU_2(\balpha_N^*;\bmx_{1i},y_i) + \bmcalI_{0}(\bmV_1^{-1}+ \bmV_{2}^{-1})^{-1}\bmV_{2}^{-1}(\hat{\balpha}_{2} - \balpha^*_{N})\notag \\ 
			&&   + \bmcalI_{0}(\bmV_1^{-1}+ \bmV_{2}^{-1})^{-1}\bmV_1^{-1}(\hat{\balpha} - \balpha^*_{N})+ o_p(n^{-1/2}),\notag \\ 
			&=& \sum_{i\in \mathcal{S}_1}\tilde{d}_{i}\bmU_2(\balpha_N^*;\bmx_{1i},y_i) + \bmcalI_{0}(\bmV_1^{-1}+ \bmV_{2}^{-1})^{-1}\bmV_{2}^{-1}(\hat{\balpha}_{2} - \balpha^*_{N})\notag \\ 
			&&  - \bmcalI_{0}(\bmV_1^{-1}+ \bmV_{2}^{-1})^{-1}\bmV_1^{-1}\bmcalI_{0}^{-1}\sum_{i\in \mathcal{S}_1}\tilde{d}_i\bmU_2(\balpha_N^*;\bmx_{1i},y_i)+ o_p(n^{-1/2}),\notag \\ 
			&=&  \bmcalI_{0}(\bmV_1^{-1}+ \bmV_{2}^{-1})^{-1}\bmV_{2}^{-1}(\hat{\balpha}_{2} - \balpha^*_{N}) +  \bmcalI_{0}(\bmV_1^{-1}+ \bmV_{2}^{-1})^{-1}\bmV_{2}^{-1}\bmcalI_{0}^{-1}\sum_{i\in \mathcal{S}_1}\tilde{d}_i\bmU_2(\balpha_N^*;\bmx_{1i},y_i)\notag \\ 
			&&+o_p(n^{-1/2})\notag \\ 
			&=&  \bmcalI_{0}\bmW(\hat{\balpha}_{2} - \balpha^*_{N}) +  \bmcalI_{0}\bmW\bmcalI_{0}^{-1}\sum_{i\in \mathcal{S}_1}\tilde{d}_i\bmU_2(\balpha_N^*;\bmx_{1i},y_i) + o_p(\kappa(n)), \label{eq: CLT alpha hat result}
		\end{eqnarray}
		where $\tilde{\balpha}$ lies on the segment joining   $\hat{\balpha}^*$ and $\balpha^*_{N}$, the second equality holds by C\ref{cond: I alpha} and Lemma~\ref{lemma: order of alpha*}, the third equality holds by (\ref{eq: alpha1 - alpha*}),  the last equality holds by C\ref{cond: W2 convergence},  $\kappa(n)=\gamma(n)$ if $\gamma(n)n^{1/2}\to\infty$ and $\kappa(n)=n^{-1/2}$ otherwise, and$\gamma(n)$ is the convergence order of $(\hat{\balpha}_{2}-\balpha^*_{N})$ in (C\ref{cond: CLT for alpha hat 2}).
		
		If there exists a non-stochastic matrix $\bmSigma_c$ such that $n\bmV_{2}=\bmSigma_c+o_p(1)$, then $(\hat{\balpha}_2 - \balpha^*_{N}) = O_p(n^{1/2})$ and $\bmW$ is not a zero matrix.  Then, by (\ref{eq: CLT alpha hat result}), we have 
		\begin{eqnarray}
			&&n^{1/2}\sum_{i\in \mathcal{S}_1}\tilde{d}_{i}
			\begin{pmatrix}
				\bmU_1(\bbeta_{0N};\bx_i,y_i)\\
				\bmU_2(\hat{\balpha}^*;\bmx_{1i},y_i)
			\end{pmatrix}\notag \\ 
			&=& \begin{pmatrix}
				\bmzero\\ n^{1/2}\bmcalI_{0}\bmW(\hat{\balpha}_{2} - \balpha^*_{N})
			\end{pmatrix} + n^{1/2}\sum_{i\in \mathcal{S}_1}\tilde{d}_{i}
			\begin{pmatrix}
				\bmU_1(\bbeta_{0N};\bmx_i,y_i)\\
				\bmcalI_{0}\bmW\bmcalI_{0}^{-1}\bmU_2(\balpha_N^*;\bmx_{1i},y_i)
			\end{pmatrix} + o_p(1).\notag \\ 
			\label{eq: dec U est}
		\end{eqnarray}
		Since the external sample is independent with the internal sample and $\bmSigma_{c}$ is the asymptotic variance of $n^{1/2}(\hat{\balpha}_{2} - \balpha^*_{N})$, by (C\ref{cond: CLT}), (C\ref{cond: CLT for alpha hat 2}) and (\ref{eq: dec U est}), we conclude that 
		\begin{eqnarray}
			n^{1/2}\sum_{i\in \mathcal{S}_1}\tilde{d}_{i}
			\begin{pmatrix}
				\bmU_1(\bbeta_{0N};\bx_i,y_i)\\
				\bmU_2(\hat{\balpha}^*;\bmx_{1i},y_i)
			\end{pmatrix}\to \mathcal{N}(0,\tilde{\bmSigma}_{u}),\notag
		\end{eqnarray}
		where  $\tilde{\bmSigma}_{11} = \bmSigma_{11}$, $\tilde{\bmSigma}_{12}=\bmSigma_{12}(\bmcalI_0^{-1})^{\top}\bmW^{\top}\bmcalI_0^{\top}$, $\tilde{\bmSigma}_{21} = \tilde{\bmSigma}_{12}^{\T}$, and $\tilde{\bmSigma}_{22} = \bmcalI_0\bmW
		\{\bmSigma_{c} +\bmcalI_0^{-1}\bmSigma_{22}(\bmcalI_0^{-1})^{\T}  \}\bmW^{\T}\bmcalI_o^{\T}$. Thus, we have proved the first case of Theorem~\ref{theo: CLT hat alpha *}.

		If $\bmW=0$, then $\gamma(n)n^{1/2}\to\infty$ and the  rate of $\kappa(n)$ is slower than $n^{-1/2}$ in (\ref{eq: CLT alpha hat result}). Thus, the remainder term of (\ref{eq: CLT alpha hat result}) is no longer $o_p(n^{-1/2})$ for $\sum_{i\in \mathcal{S}_1}\tilde{d}_{i}\bmU_2(\hat{\balpha}^*;\bmx_{1i},y_i)$. Instead, for this case, we investigate the asymptotic order of $(\bmV_1^{-1}+ \bmV_{2}^{-1})^{-1}\bmV_{2}^{-1}$ first. By C\ref{cond: CLT}, C\ref{cond: var consist alphas} and (\ref{eq: alpha1 - alpha*}), we have \begin{equation}
			\bmV_1 \asymp n^{-1}\quad \mbox{and}\quad \bmV_{2}\asymp \gamma(n)^{-2}\label{eq: asympt order V1 V2}
		\end{equation}
		in probability. Thus, (\ref{eq: asympt order V1 V2}) leads to 
		\begin{equation}
			(\bmV_1^{-1}+ \bmV_{2}^{-1})^{-1}\bmV_{2}^{-1}\asymp n^{-1}\gamma(n)^{2} \label{eq: V conv rate}
		\end{equation}
		in probability by the fact that $\gamma(n)n^{1/2}\to\infty$. Thus, by (C\ref{cond: CLT for alpha hat 2}) and (\ref{eq: V conv rate}), we have 
		\begin{equation}
			(\bmV_1^{-1}+ \bmV_{2}^{-1})^{-1}\bmV_{2}^{-1}(\hat{\balpha}_{2} - \balpha^*_{N}) \asymp n^{-1}\gamma(n)\label{eq: first order}
		\end{equation}
		in probability. By $\gamma(n)n^{1/2}\to\infty$, (\ref{eq: V conv rate}) and (\ref{eq: first order}), we have shown 
		\begin{eqnarray}
			&(\bmV_1^{-1}+ \bmV_{2}^{-1})^{-1}\bmV_{2}^{-1} = o_p(1)\label{eq: first part vv inv}\notag,\\
			&(\bmV_1^{-1}+ \bmV_{2}^{-1})^{-1}\bmV_{2}^{-1}(\hat{\balpha}_{2} - \balpha^*_{N})  = o_p(n^{-1/2})\notag
		\end{eqnarray}
		in probability. Thus, by the fourth equality of (\ref{eq: CLT alpha hat result}), we can show that 
		$$
		\sum_{i\in \mathcal{S}_1}\tilde{d}_{i}\bmU_2(\hat{\balpha}^*;\bmx_{1i},y_i) = o_p(n^{-1/2}),
		$$
		and we have proved the third case of Theorem~\ref{theo: CLT hat alpha *}.

	\end{proof}
	
\section*{Supplementary Material}
	\renewcommand{\theequation}{S.\arabic{equation}}
	\renewcommand{\thelemma}{S\arabic{lemma}}
	\renewcommand{\thetheorem}{S\arabic{theorem}}
	\renewcommand{\thetable}{S\arabic{table}}
    \renewcommand{\thefigure}{S\arabic{figure}}
	\setcounter{equation}{0}
	\setcounter{theorem}{0}
	\setcounter{lemma}{0}
	\setcounter{section}{0}

	\renewcommand{\thesection}{S\arabic{section}}	
		\section{Special case under simple random sampling}\label{SS: hat beta variance}
	By  C\ref{cond: I 22}  and C\ref{cond: variance conv}, both $\bmcalI_{11}$ and $\bmcalI_{22}$  are invertible, so we have 
	\begin{equation}
		\bmcalI^{-1} 
		= \begin{pmatrix}
			\bmcalI_{11}^{-1}&-\bmcalI_{11}^{-1}\bmcalI_{12}\bmcalI_{22}^{-1}\\
			\bmzero&\bmcalI_{22}^{-1}
		\end{pmatrix}.\label{eq: I inv}
	\end{equation}
	By (\ref{eq: I inv}), it leads to 
	\begin{eqnarray}
		\bmcalI^{-1} \bmSigma_u 
		&=& \begin{pmatrix}
			\bmcalI_{11}^{-1}&-\bmcalI_{11}^{-1}\bmcalI_{12}\bmcalI_{22}^{-1}\\
			\bmzero&\bmcalI_{22}^{-1}
		\end{pmatrix}\begin{pmatrix}
			\bmSigma_{11}&\bmSigma_{12}\\
			\bmSigma_{21}&\bmSigma_{22}
		\end{pmatrix}\notag \\ 
		&=&\begin{pmatrix}
			\bmcalI_{11}^{-1}\bmSigma_{11} - \bmcalI_{11}^{-1}\bmcalI_{12}\bmcalI_{22}^{-1}\bmSigma_{21} & \bmcalI_{11}^{-1}\bmSigma_{12} - \bmcalI_{11}^{-1}\bmcalI_{12}\bmcalI_{22}^{-1}\bmSigma_{22}\\
			\bmcalI_{22}^{-1}\bmSigma_{21}& \bmcalI_{22}^{-1}\bmSigma_{22}
		\end{pmatrix}.\label{eq: I SIGMA}
	\end{eqnarray}
	By (\ref{eq: I inv})--(\ref{eq: I SIGMA}), we have 
	\begin{eqnarray}
		&&\bmcalI^{-1} \bmSigma_u(\bmcalI^{-1})^{\T}\notag \\ 
		&=&\begin{pmatrix}
			\bmcalI_{11}^{-1}\bmSigma_{11} - \bmcalI_{11}^{-1}\bmcalI_{12}\bmcalI_{22}^{-1}\bmSigma_{21} & \bmcalI_{11}^{-1}\bmSigma_{12} - \bmcalI_{11}^{-1}\bmcalI_{12}\bmcalI_{22}^{-1}\bmSigma_{22}\\
			\bmcalI_{22}^{-1}\bmSigma_{21}& \bmcalI_{22}^{-1}\bmSigma_{22}
		\end{pmatrix}\begin{pmatrix}
			(\bmcalI_{11}^{-1})^{\top}&\bmzero\\
			-\bmcalI_{22}^{-1}\bmcalI_{12}^{\T}(\bmcalI_{11}^{-1})^{\T}&\bmcalI_{22}^{-1}
		\end{pmatrix}\notag \\ 
		&=& \begin{pmatrix}
			\bmA&\bmB\\
			\bmC&\bmD
		\end{pmatrix},\notag
	\end{eqnarray}
	where 
	\begin{eqnarray}
		\bmA &=& \bmcalI_{11}^{-1}\bmSigma_{11}(\bmcalI_{11}^{-1})^{\T} -  \bmcalI_{11}^{-1}\bmcalI_{12}\bmcalI_{22}^{-1} \bmSigma_{21}(\bmcalI_{11}^{-1})^{\T} - \bmcalI_{11}^{-1}\bmSigma_{12}\bmcalI_{22}^{-1} \bmcalI_{12}^{\T}(\bmcalI_{11}^{-1})^{\T}\notag \\ 
		&&+\bmcalI_{11}^{-1}\bmcalI_{12}\bmcalI_{22}^{-1} \bmSigma_{22}\bmcalI_{22}^{-1} \bmcalI_{12}^{\T}(\bmcalI_{11}^{-1})^{\T},\notag \\
		\bmB  &=& \bmcalI_{11}^{-1}\bmSigma_{12} \bmcalI_{22}^{-1} - \bmcalI_{11}^{-1}\bmcalI_{12}\bmcalI_{22}^{-1} \bmSigma_{22} \bmcalI_{22}^{-1},\notag \\ 
		\bmC &=& \bmcalI_{22}^{-1}\bmSigma_{21}(\bmcalI_{11}^{-1})^{\T}-\bmcalI_{22}^{-1}\bmSigma_{22}\bmcalI_{22}^{-1}\bmcalI_{12}^{\T}(\bmcalI_{11}^{-1})^{\T},\notag \\ 
		\bmD &=& \bmcalI_{22}^{-1}\bmSigma_{22}\bmcalI_{22}^{-1},\notag\\
		\label{eq: D}
	\end{eqnarray}
	and $\bmA$ in (\ref{eq: D}) is the asymptotic variance of $n^{1/2}(\hat{\bbeta}-\bbeta_{0,N})$. 
	
	Next, consider simple random sampling without replacement under the assumption  $nN^{-1}\to0$, so the sampling weight is $d_i = Nn^{-1}$ under such a design. Besides, we have  
	\begin{eqnarray}
		&&\var\left\{ n^{1/2}N^{-1}\sum_{i\in \mathcal{S}_1} \begin{pmatrix}
			\bmU(\bbeta_{0N};\bx_i,y_i)\\
			\bmU(\balpha_N^*;\bmx_{1i},y_i)
		\end{pmatrix}\right\}= (1-nN^{-1})(N-1)^{-1}\notag \\ 
		% &=& (N-1)^{-1}(1 - nN^{-1})\times\notag \\ 
		% &&
		% \begin{pmatrix}
		%             \displaystyle \sum_{i=1}^N\bmU(\balpha_N^*;\bmx_{1i},y_i)^{\otimes2} &\displaystyle \sum_{i=1}^N \bmU(\balpha_N^*;\bmx_{1i},y_i)\left\{\bmU(\bbeta_{0N};\bx_i,y_i) - \bar{\bmU}_N(\bmbeta_{0N})\right\}^{\T}\\
		%             \displaystyle  \sum_{i=1}^N \left\{\bmU(\bbeta_{0N};\bx_i,y_i) - \bar{\bmU}_N(\bmbeta_{0N})\right\}\bmU(\balpha_N^*;\bmx_{1i},y_i)^{\T} &  \displaystyle \sum_{i=1}^N \left\{\bmU(\bbeta_{0N};\bx_i,y_i) - \bar{\bmU}_N(\bmbeta_{0N})\right\}^{\otimes2}
		% \end{pmatrix}\notag \\
		&&\times\begin{pmatrix}
			\displaystyle \sum_{i=1}^N \bmU(\bbeta_{0N};\bx_i,y_i)^{\otimes2} &\displaystyle  \sum_{i=1}^N \bmU(\bbeta_{0N};\bx_i,y_i) \bmU(\balpha_N^*;\bmx_{1i},y_i)^{\T}\\
			\displaystyle \sum_{i=1}^N \bmU(\balpha_N^*;\bmx_{1i},y_i)\bmU(\bbeta_{0N};\bx_i,y_i) ^{\T} & \displaystyle \sum_{i=1}^N \bmU(\balpha_N^*;\bmx_{1i},y_i)^{\otimes2} 
		\end{pmatrix}\notag
	\end{eqnarray}
	 where the equality holds since $\sum_{i=1}^N\bmU(\balpha_N^*,x_{1i},y_{1i})=\bmzero$ and $\sum_{i=1}^N\bmU(\bbeta_{0N};\bx_i,y_i)=\bmzero$. Since the sampling fraction is asymptotically negligible, by (C\ref{cond: CLT}), we conclude that 
	\begin{eqnarray}
		&\bmSigma_{11} = N^{-1} \sum_{i=1}^N \bmU(\bbeta_{0N};\bx_i,y_i)^{\otimes2}  + o_p(1),\notag \\
		&\bmSigma_{12} = \bmSigma_{21}^{\T} = N^{-1} \sum_{i=1}^N\bmU(\bbeta_{0N};\bx_i,y_i)  \bmU(\balpha_N^*;\bmx_{1i},y_i)^{\T}  + o_p(1),\label{eq: S sigma 12 srs}\\
		&\bmSigma_{22} =  N^{-1}  \sum_{i=1}^N \bmU(\balpha_N^*;\bmx_{1i},y_i)^{\otimes2}  + o_p(1)\label{eq: S sigma 11 srs}
	\end{eqnarray}  Besides, by (C\ref{cond: variance conv})--(C\ref{cond: I 22}) and the basic theoretical properties of simple random sampling without replacement, we can also get 
	\begin{eqnarray}
		\bmcalI_{12} &=&\sum_{i\in \mathcal{S}_1}\tilde{d}_i \bmU(\bbeta_{0N};\bmx_{i},y_i)\bmU(\balpha_N^*;\bmx_{1i},y_i)^{\T} + o_p(1) \notag \\ 
		&=&  N^{-1} \sum_{i=1}^N\bmU(\bbeta_{0N};\bmx_i,y_i)  \bmU(\balpha_N^*;\bmx_{1i},y_i)^{\T}  + o_p(1),\label{eq: S I 21 srs}\\
		\bmcalI_{22} &=& \sum_{i\in \mathcal{S}_1}\tilde{d}_i\bmU(\balpha_N^*;\bmx_{1i},y_i)^{\otimes2} + o_p(1)\notag \\
		&=& N^{-1} \sum_{i=1}^N \bmU(\balpha_N^*;\bmx_{1i},y_i)^{\otimes2}  + o_p(1),\label{eq: S I 11 srs}
	\end{eqnarray}
	where $\tilde{d}_i=n^{-1}$ under simple random sampling without replacement.
	
	By (\ref{eq: S sigma 12 srs})--(\ref{eq: S I 11 srs}), we conclude that 
	\begin{equation}
		\bmSigma_{12}=\bmcalI_{12},\quad \bmSigma_{22} = \bmcalI_{22} \label{eq: S equv I S srs}
	\end{equation}
	under simple random sampling without replacement. Then, by (\ref{eq: S equv I S srs}), the asymptotic variance of  $n^{1/2}(\hat{\bbeta}-\bbeta_0)$ can be simplified as
	\begin{eqnarray}
		\bmA &=& \bmcalI_{11}^{-1}\bmSigma_{11}(\bmcalI_{11}^{-1})^{\T} -  \bmcalI_{11}^{-1}\bmcalI_{12}\bmcalI_{22}^{-1} \bmSigma_{21}(\bmcalI_{11}^{-1})^{\T} - \bmcalI_{11}^{-1}\bmSigma_{12}\bmcalI_{22}^{-1} \bmcalI_{12}^{\T}(\bmcalI_{11}^{-1})^{\T}\notag \\ 
		&&+\bmcalI_{11}^{-1}\bmcalI_{12}\bmcalI_{22}^{-1} \bmSigma_{22}\bmcalI_{22}^{-1} \bmcalI_{12}^{\T}(\bmcalI_{11}^{-1})^{\T},\notag \\
		&=& \bmcalI_{11}^{-1}\bmSigma_{11}(\bmcalI_{11}^{-1})^{\T} -  \bmcalI_{11}^{-1}\bmSigma_{12}\bmSigma_{22}^{-1} \bmSigma_{21}(\bmcalI_{11}^{-1})^{\T}\notag \\ 
		&=& \bmcalI_{11}^{-1}(\bmSigma_{11} - \bmSigma_{12}\bmSigma_{22}^{-1} \bmSigma_{21})(\bmcalI_{11}^{-1})^{\T}.\label{eq: srs general result}\notag
	\end{eqnarray}
	Thus, the proposed working model approach improves the estimation efficient of $\hat{\bbeta}$ under simple random sampling without replacement. We can draw a similar conclusion for simple random sampling with replacement, and we do not need to assume $nN^{-1}\to0$ under such a design.
	
		\section{Implementation of \citet{chatterjee2016}}\label{sec: super-popu  model}
	Assume that the finite population $\{(\bmx_i,y_i):i=1,\ldots,N\}$ is a random sample from a super-population model with conditional density $f(y\mid \bx;\btheta_f)$ with parameter $\btheta_f$. Refer $g(y\mid \bx_1;\btheta_r)$ as the ``reduced'' model with parameter $\btheta_r$. For simplicity, we assume that the intercept term is included in $\bx$. In this section, the parameters are denoted as $\btheta_f$ and $\btheta_r$, and we use $\bmbeta$ and $\bmalpha$ as the regression coefficients.
	\subsection{Linear regression model}\label{ss: linear CML}
	Assume that $f(y\mid \bx;\btheta_f) = (2\pi\sigmaot)^{-1/2}\exp\{-(y-\bx^{\T}\bbeta)^2/(2\sigmaot)\}$ corresponds to a normal density function with $\btheta_f = (\bbeta^{\T},\sigmaot)^{\T}$, and  assume  another normal density function $g(y\mid \bx_1;\btheta_r) = (2\pi\sigmatt)^{-1/2}\exp\{-(y-\bz^{\T}\balpha)^2/(2\sigmatt)\}$ for the reduced model, where $\bz = (1,\bmx_{1}^{\T})^{\T}$, and $\btheta_r = (\balpha^{\T},\sigmatt)^{\T}$. Assume that $\btheta_r$ is available, and denote $\beeta = (\blambda^{\T},\btheta_f^{\T})^{\T}$,
	\begin{eqnarray}
		&\bs_{\beta}(y_i,\bxi;\beeta ) = \begin{pmatrix}
			\ddfrac{y_i-\bxi^{\T}\bbeta}{\sigmaot}\bxi\\[0.5cm]
			-\ddfrac{1}{2\sigmaot} + \ddfrac{(y_i-\bxi^{\T}\bbeta)^2}{2(\sigmaot)^2}
		\end{pmatrix}, \quad \bu_{\beta}(\bxi;\beeta ) = \begin{pmatrix}
			\ddfrac{\bxi^{\T}\bbeta - \bxoi^{\T}\balpha}{(\sigmatt)^2}\bxoi\\[0.5cm]
			-\ddfrac{1}{2\sigmatt} + \ddfrac{\sigmaot + (\bxi^{\T}\bbeta - \bxoi^{\T}\balpha)^2}{2(\sigmatt)^2}
		\end{pmatrix},\notag 
	\end{eqnarray}
	\begin{eqnarray}
		& \bc_{\beta}(\bxi;\beeta ) = \begin{pmatrix}
			\ddfrac{\bxi\bxoi^{\T}}{\sigmatt}&\ddfrac{\bxi^{\T}\bbeta - \bxoi^{\T}\balpha}{(\sigmatt)^2}\bxi\\[0.5cm]
			0&\ddfrac{1}{2(\sigmatt)^2}
		\end{pmatrix},\quad \tilde{\bs}_{\beta}(\bxi;\beeta) = \ddfrac{\bc_{\beta}(\bxi;\beeta)\blambda}{1-\blambda^{\T}\bu_{\beta}(\bxi;\beeta)},\notag
	\end{eqnarray}
	\begin{eqnarray}
		& \bs^*_{\beta}(y_i,\bxi;\beeta ) = \bs_{\beta}(y_i,\bxi;\beeta) + \tilde{\bs}_{\beta}(\bxi;\beeta),\quad \bs^*_{\lambda}(\bxi;\beeta) = \ddfrac{ \bu_{\beta}(\bxi;\beeta )}{1-\blambda^{\T}\bu_{\beta}(\bxi;\beeta)}.\notag
	\end{eqnarray}
	Then, we are interested in solving 
	\begin{equation}
		\bg^*(\beeta) = \begin{pmatrix}
			\displaystyle\sum_{i\in \mathcal{S}_1} \bs^*_{\beta}(y_i,\bxi;\beeta )\\[0.5cm]
			\displaystyle\sum_{i\in \mathcal{S}_1} \bs^*_{\lambda}(\bxi;\beeta )
		\end{pmatrix}=\bmzero.\label{eq: target}
	\end{equation}
	We use a modified Newton-Raphson algorithm  \citep{wu2005algorithms} to solve (\ref{eq: target}). Denote $\bI^*(\beeta) = -\partial\bg^*(\beeta)/\partial \beeta^{\T}$,
	\begin{eqnarray}
		\bi_{\beta\beta}(y_i,\bxi;\beeta) = \begin{pmatrix}
			\ddfrac{\bxi\bxi^{\T}}{\sigmaot}&\ddfrac{y_i-\bxi^{\T}\bbeta}{(\sigmaot)^2}\bxi\\[0.5cm]
			\ddfrac{y_i-\bxi^{\T}\bbeta}{(\sigmaot)^2}\bxi^{\T}&\ddfrac{(y_i-\bxi^{\T}\bbeta)^2}{(\sigmaot)^3} - \ddfrac{1}{2(\sigmaot)^2}
		\end{pmatrix},\quad \bd_{\beta}(\bxi;\beeta) = \begin{pmatrix}
			-\ddfrac{\lambda_3\bxi\bxi^{\T}}{(\sigmatt)^2}&\bmzero\\[0.5cm]
			\bmzero^\top&0
		\end{pmatrix}, \notag
	\end{eqnarray}
	\begin{eqnarray}
		&\bI^*_{\beta\beta}(\beeta) = \sum_{i\in \mathcal{S}_1}\bi_{\beta\beta}(y_i,\bxi;\beeta)  + \ddfrac{\bd_{\beta}(\bxi;\beeta)}{1-\blambda^{\T}\bu_{\beta}(\bxi;\beeta)} - \tilde{\bs}_{\beta}(\bxi;\beeta)^{\otimes2}, \notag\\
		&\bI^*_{\beta\lambda}(\beeta)=\{\bI^*_{\lambda\beta}(\beeta)\}^{\T} = -\sum_{i\in \mathcal{S}_1}\left[\ddfrac{\bc_{\beta}(\bxi;\beeta ) }{1-\blambda^{\T}\bu_{\beta}(\bxi;\beeta)} + \tilde{\bs}_{\beta}(\bxi;\beeta) \{\bs^*_{\lambda}(\bxi;\beeta)\}^{\T}\right],\notag \\ 
		&\bI^*_{\lambda\lambda}(\beeta) = -\sum_{i\in S}\bs^*_{\lambda}(\bxi;\beeta)^{\otimes2},\notag
	\end{eqnarray}
	where the corresponding components of $\bd_{\beta}(\bxi;\beeta)$ has the same dimension as that of $ \bi_{\beta\beta}(y_i,\bxi;\beeta)$. Then, 
	\begin{equation}
		\bI^*(\beeta) = \begin{pmatrix}
			\bI^*_{\beta\beta}(\beeta) & \bI^*_{\beta\lambda}(\beeta)\\[0.5cm]
			\bI^*_{\lambda\beta}(\beeta) &\bI^*_{\lambda\lambda}(\beeta) 
		\end{pmatrix}\notag
	\end{equation}
	
	Denote $\btheta_f^{(0)}=\hat{\btheta}_f$ and  $\blambda^{(0)} = (0,0,0)^{\T}$, where $\hat{\btheta}_f$ is a design-based estimator using the probability sample $\mathcal{S}_1$. The following is the modified Newton-Raphson method. 
	\begin{enumerate}
		\item Initialize $\beeta^{(0)} = (\blambda^{(0)},\btheta_f^{(0)})$.
		\item For the $k$th iteration, 
		\begin{enumerate}
			\item Obtain $\bg^{*(k+1)} = \bg^*(\beeta^{(k)})$,
			\item Obtain $\bI^{*(k+1)} =\bI^*(\beeta^{(k)}) $,
			\item Obtain $\bbbdelta_t = (\bI^{*(k+1)} )^{-1}\bg^{*(k+1)}$,\label{enu: d}
			\item Obtain $\beeta_t = \beeta^{(k)} + \bbbdelta_t$,
			\item If $\min\{1-\blambda_t^{\T}\bu_{\beta}(\bxi;\beeta_t):i\in \mathcal{S}_1\}<0$ or the number of iterations is less than a threshold, set $\bbbdelta_t = \bbbdelta_t/2$ and go back to (\ref{enu: d}), where $\blambda_t$ is the corresponding component of $\beeta_t$.
			\item If $\min\{1-\blambda_t^{\T}\bu_{\beta}(\bxi;\beeta_t):i\in \mathcal{S}_1\}>0$, set $\beeta^{(k+1)} = \beeta^{(k)} + \bbbdelta_t$,
			\item If $\min\{1-\blambda_t^{\T}\bu_{\beta}(\bxi;\beeta_t):i\in\mathcal{S}_1\}<0$, break all the iterations and return NA.
		\end{enumerate}
		\item Go back to Step 2 until convergence. If the number of iteration reaches a threshold, then return NA.
	\end{enumerate}
	\subsection{Logistic regression model}\label{ss: lrm}
	When the response of interest is binary, we consider the following full model:
	\begin{equation}
		\mbox{logit}\{\Pr(Y=1\mid \bx;\bbeta)\} = \bx^{\T}\bbeta,\notag
	\end{equation}
	where $\mbox{logit}(p) = \log(p) - \log(1-p)$ for $p\in(0,1)$. Besides, we consider the following reduced model:
	\begin{equation}
		\mbox{logit}\{\Pr(Y=1\mid \bz;\balpha)\} = \bz^{\T}\balpha,\notag
	\end{equation}
	where $\bz$ contains the covariates for the reduced model. 
	
	Denote $\beeta = (\blambda^{\T},\bbeta^{\T})^{\T}$, and we have 
	\begin{eqnarray}
		&\bs_{\beta}(y_i,\bxi;\beeta ) = \{y_i - p_i(\bxi;\bbeta)\}\bxi, \quad \bu_{\beta}(\bxi;\beeta ) =\{p_i(\bxi;\bbeta)-p_{1i}(\bxoi;\balpha) \}\bxoi,\notag \\
		& \bc_{\beta}(\bxi;\beeta ) = p_i(\bxi;\bbeta)\{1-p_i(\bxi;\bbeta)\}\bxi\bxoi^{\T},\quad \tilde{\bs}_{\beta}(\bxi;\beeta) = \ddfrac{\bc_{\beta}(\bxi;\beeta)\blambda}{1-\blambda^{\T}\bu_{\beta}(\bxi;\beeta)},\notag\\
		&\bi_{\beta\beta}(y_i,\bxi;\beeta) =p_i(\bxi;\bbeta) \{1-p_i(\bxi;\bbeta)  \}\bxi\bxi^{\T},\notag\\
		& \bd_{\beta}(\bxi;\beeta) =  \blambda^{\T}\bxoi p_i(\bxi;\bbeta) \{1-p_i(\bxi;\bbeta)  \}\{2p_i(\bxi;\bbeta)-1\}\bxi\bxi^{\T},\notag
	\end{eqnarray}
	where $ p_i(\bxi;\bbeta) = \Pr(Y_i=1\mid \bxi;\bbeta)$, $ p_{1i}(\bxoi;\balpha) = \Pr(Y_i=1\mid \bxoi;\balpha)$, and $\bxoi = (1,\bmx_{1i}^{\T})^{\T}$. Then, we can use the same procedure to estimate the the corresponding parameters.
	\section{Additional simulation study}\label{ss: logistic regression}
	The additional   simulation study assumes that the response of interest is a binary outcome. The covariates $\bmx_i=(x_{i1},x_{i2})^\top$  are generated by the same setups in the previous section. Then, $y_i$ is generated by a Bernoulli distribution with success probability $\Pr(Y_i = 1\mid  x_{i1}, x_{i2}) = \text{logit}^{-1}( \beta_0 +\beta_1 x_{i1} + \beta_2 x_{i2})$ with the simulation parameters $(\beta_0,\beta_1,\beta_2) = (-0.5,0.3,-0.1)$. 
	We consider two sampling schemes to generate a probability sample $\mathcal{S}_1$ of size $n_1=1{,}000$: (i) SRS and (ii) Poisson sampling with inclusion probabilities satisfying $\pi_i \propto 0.9I(y_i=1) + 0.1I(y_i=0)$ and $\sum_{i=1}^N\pi_i=n_1$.
	% Two different sample sizes are considered: $(n_1,n_2) = (200,5{,}000)$ and $(n_1,n_2) = (5{,}000,50{,}000)$. 
	
	For the proposed estimator, we consider a working reduced model, $\bmU_2(\bmalpha;x_{i1},y_i)=\{y_i - \text{expit}(\alpha_0+\alpha_1 x_{i1})\} (1,x_{i1})^\top$, where $\text{expit}(x) = \{1+\exp(-x)\}^{-1}$. 
	Similar to the first simulation,  we consider two sampling designs to generate an external sample $\mathcal{S}_2$ of (expected) size $n_2 = 10{,}000$: (i) SRS and (ii) Poisson sampling with inclusion probabilities satisfying $\pi_{2i}\propto\{1 + \exp(0.2x_{i1} + 0.1x_{i2}-0.6)\}^{-1}$ and $\sum_{i=1}^N\pi_{2i}=n_2$. We still compare the two estimators in the first simulation; see \ref{ss: lrm} of the Supporting Information for details about the CML estimator.
	
	% To estimate the parameters for the logistic regression model, we consider the following method. 
	% \begin{enumerate}
	%     \item[Pr1] The proposed method, and the parameters in the reduced model is estimated based on the probability sample $S$ and an external sample $S_e$ generated by simple random sampling with sample size $n_2$.
	%     \item[Ch1] The \citet{chatterjee2016}'s estimator using the estimated reduced model parameters from method Pr1.
	%     \item[Pr2] The proposed method, and the parameters in the reduced model is estimated based on the finite population.
	%     \item[Ch2] The \citet{chatterjee2016}'s estimator using the estimated reduced model parameters from method Pr2.
	%     \item[PS] The parameters in the full model are estimated by the probability sample $S$ only.
	% \end{enumerate}
	
\begin{figure}[!ht]
		\centering %This figure is generated by Second_simulation_Call_v1.R and Second_simulation_Call_v1.R in the server
		\includegraphics[width=\textwidth]{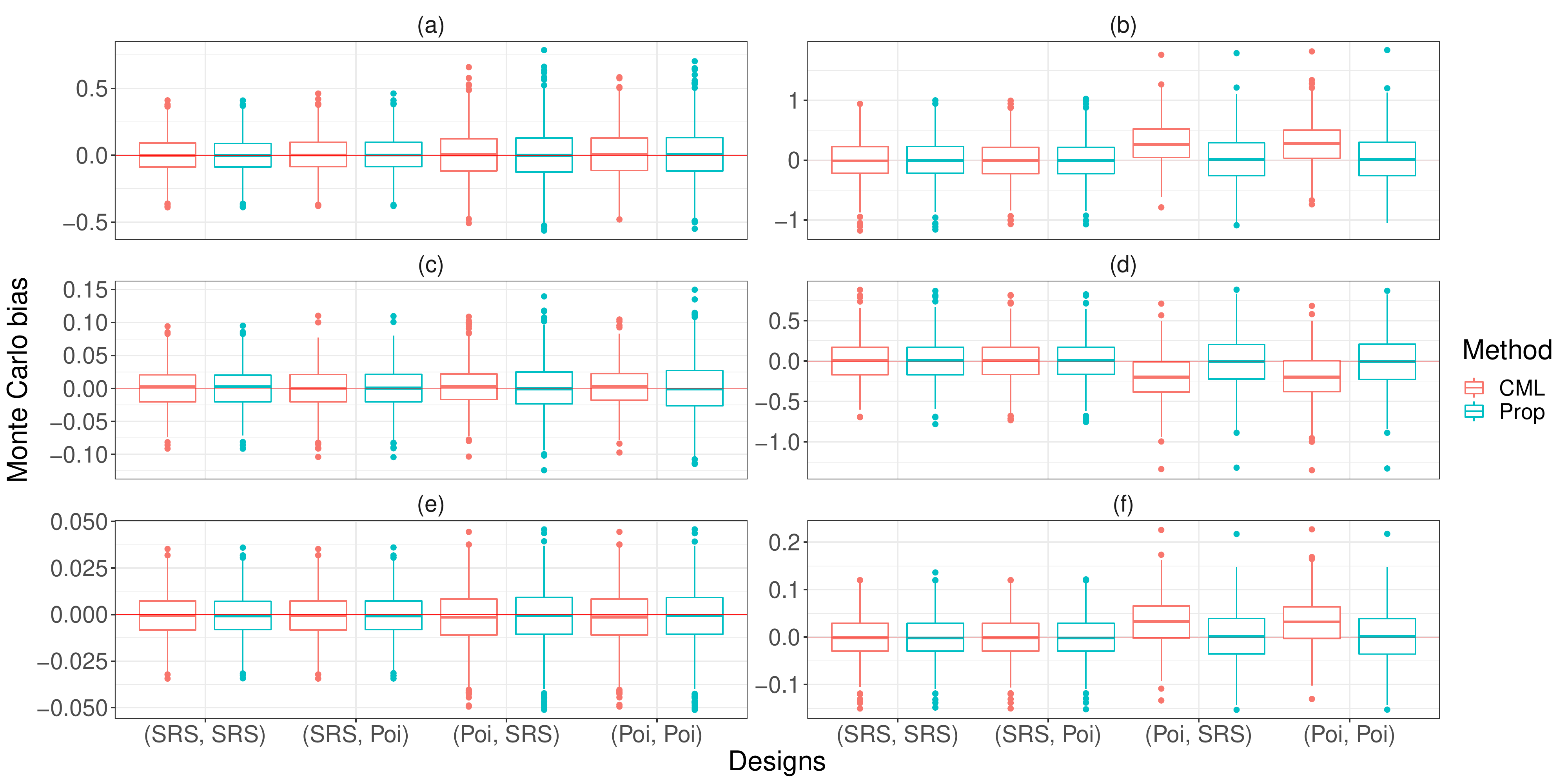}
		\caption{Monte Carlo bias of   the proposed and CML estimators  based on 1,000 Monte Carlo simulations under the logistic regression model setup. The first to the third rows stand for the Monte Carlo bias for estimating $\beta_0$, $\beta_1$ and $\beta_2$, respectively. The three plots in the left column show the results when the auxiliary variables are independently generated, and those in the right column are for the case when the auxiliaries are dependent.  ``CML'' and ``Prop'' stands for the CML estimator and the proposed estimator, respectively. The first design in the parenthesis is used to generate the internal sample $\mathcal{S}_1$, and the second one to generate the external sample $\mathcal{S}_2$. ``Poi'' represents Poisson sampling.}
		\label{fig: lrg}
	\end{figure}

	We conduct $M=1{,}000$ Monte Carlo simulations, and Web Figure~\ref{fig: lrg} shows the  Monte Carlo bias of  the proposed and CML estimators, and we can observe similar patterns as in the first simulation study. When the covariates are independent, both methods perform approximately the same. However, when the covariates are dependent, the CML method leads to biased estimators when the internal sample $\mathcal{S}_1$ is generated by an informative Poisson sampling design.
	
	Web Table~\ref{tab: LRM} shows the coverage rate of a 95\% confidence intervals  for the proposed estimator under different settings. As in the first simulation, the coverage rats are close to their nominal truth 0.95 under different settings, indicating the satisfactory performance of the proposed estimator.
	\begin{table}[!ht]
		% Generated by Second_simulation_Call_v1.R on the department cluster
		\footnotesize
		\renewcommand{\arraystretch}{0.75}
		\centering% This table is typed based the previous result.
		\caption{Coverage rate of a 95\% confidence interval by the proposed method based on 1,000 Monte Carlo simulations under different setups. ``$\mathcal{S}_1$ Des'' and ``$\mathcal{S}_2$ Des'' show the sampling design used to generate the internal sample $\mathcal{S}_1$  and the external sample $\mathcal{S}_2$. ``SRS'' and ``Poi'' stands for SRS and Poisson sampling, respectively. ``Independent'' and ``Dependent'' correspond to the cases when the auxiliary variables are independent and dependent, respectively.  }
		\begin{tabular}{ccccccccc}
			\hline
			\multirow{2}{*}{$\mathcal{S}_1$ Des}&\multirow{2}{*}{$\mathcal{S}_2$ Des}&\multicolumn{3}{c}{Independent}&&\multicolumn{3}{c}{Dependent}\\
			\cline{3-5}\cline{7-9}
			&&$\beta_0$&$\beta_1$&$\beta_2$&&$\beta_0$&$\beta_1$&$\beta_2$\\
			\hline
			\multirow{2}{*}{SRS}&SRS &0.963&0.952&0.959&&0.957&0.953&0.949\\
			&Poi &0.948&0.955&0.959&&0.953&0.954&0.950\\\vspace{-1em}\\
			\multirow{2}{*}{Poi}&SRS &0.954&0.949&0.940&&0.951&0.949&0.946\\
			&Poi &0.941&0.952&0.940&&0.952&0.950&0.948\\
			\hline                          
		\end{tabular}
		
		\label{tab: LRM}
	\end{table}
\section{Validation for the linearity assumption for the KNHANES dataset}\label{ss: valid linearity}
For the demonstration purpose, we have assumed two linear regression models in Section~\ref{sec:application_result}. In this section, we validate the linearity assumption for the regression model.

\begin{figure}[!h]
    \centering
    \includegraphics[width=.5\textwidth]{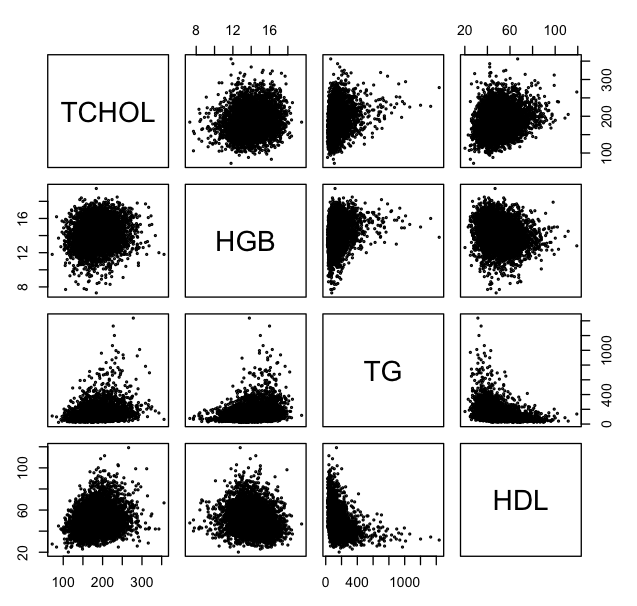}
    \caption{Scatter plots for ``Total Cholesterol (TCHOL)'', ``Hemoglobin (HGB)'', ``Triglyceride (TG)'' and ``HDL''. }
    \label{fig: app linear}
\end{figure}

Figure~\ref{fig: app linear} shows the relationship among the response of interest ``Total Cholesterol'' and the three covariates. We can conclude that the proposed two linear regression models are reasonable to analyze the KNHANES dataset. Please notice that the KNHANES dataset demonstrates heterogeneity for the linear regression model. Such heterogeneity only influences the efficiency of the proposed method, and it does not invalidate the linear regression model.

\bibliographystyle{chicago}
\bibliography{ref}

\end{document}